\documentclass[twocolumn,showpacs,aps,prd,floatfix]{revtex4}
\usepackage{graphicx}
\usepackage{overpic}
\usepackage[usenames]{xcolor}
\usepackage{relsize}
\usepackage{rotating}
\usepackage{dcolumn}
\usepackage{amsmath}
\usepackage{epsfig}
\usepackage{multirow}
\usepackage{subfigure}

\newcommand{\BABARPubYear}{15}
\newcommand{\BABARPubNumber}{002}
\newcommand{\SLACPubNumber}{16384}

\RequirePackage{xspace}

\usepackage{relsize}
\def\babar{\mbox{\slshape B\kern-0.1em{\smaller A}\kern-0.1em
    B\kern-0.1em{\smaller A\kern-0.2em R}}}

\def\epem       {\ensuremath{e^+e^-}\xspace}

\def\mumu       {\ensuremath{\mu^+\mu^-}\xspace}

\def\ellell     {\ensuremath{\ell^+ \ell^-}\xspace}

\def\qqbar {\ensuremath{q\overline q}\xspace}

\def\ccbar {\ensuremath{c\overline c}\xspace}

\def\piz   {\ensuremath{\pi^0}\xspace}

\def\pip   {\ensuremath{\pi^+}\xspace}
\def\pim   {\ensuremath{\pi^-}\xspace}

\def\kaon  {\ensuremath{K}\xspace}
\def\Kbar  {\kern 0.2em\overline{\kern -0.2em K}{}\xspace}

\def\Kz    {\ensuremath{K^0}\xspace}
\def\Kzb   {\ensuremath{\Kbar^0}\xspace}
\def\KzKzb {\ensuremath{\Kz \kern -0.16em \Kzb}\xspace}
\def\Kp    {\ensuremath{K^+}\xspace}
\def\Km    {\ensuremath{K^-}\xspace}

\def\KpKm  {\ensuremath{\Kp \kern -0.16em \Km}\xspace}
\def\KS    {\ensuremath{K^0_{\scriptscriptstyle S}}\xspace}

\def\Kstarz  {\ensuremath{K^{*0}}\xspace}

\def\Kstar   {\ensuremath{K^*}\xspace}
\def\Kstarb  {\ensuremath{\Kbar^*}\xspace}
\def\Kstarp  {\ensuremath{K^{*+}}\xspace}

\def\Dbar    {\kern 0.2em\overline{\kern -0.2em D}{}\xspace}

\def\Dz      {\ensuremath{D^0}\xspace}
\def\Dzb     {\ensuremath{\Dbar^0}\xspace}
\def\DzDzb   {\ensuremath{\Dz {\kern -0.16em \Dzb}}\xspace}
\def\Dp      {\ensuremath{D^+}\xspace}
\def\Dm      {\ensuremath{D^-}\xspace}

\def\DpDm    {\ensuremath{\Dp {\kern -0.16em \Dm}}\xspace}

\def\B       {\ensuremath{B}\xspace}
\def\Bbar    {\kern 0.18em\overline{\kern -0.18em B}{}\xspace}

\def\BB      {\ensuremath{B\Bbar}\xspace}
\def\Bz      {\ensuremath{B^0}\xspace}
\def\Bzb     {\ensuremath{\Bbar^0}\xspace}
\def\BzBzb   {\ensuremath{\Bz {\kern -0.16em \Bzb}}\xspace}
\def\Bu      {\ensuremath{B^+}\xspace}
\def\Bub     {\ensuremath{B^-}\xspace}
\def\Bp      {\ensuremath{\Bu}\xspace}

\def\BpBm    {\ensuremath{\Bu {\kern -0.16em \Bub}}\xspace}

\def\BorBbar    {\kern 0.18em\optbar{\kern -0.18em B}{}\xspace}
\def\DorDbar    {\kern 0.18em\optbar{\kern -0.18em D}{}\xspace}
\def\KorKbar    {\kern 0.18em\optbar{\kern -0.18em K}{}\xspace}

\def\jpsi     {\ensuremath{{J\mskip -3mu/\mskip -2mu\psi\mskip 2mu}}\xspace}
\def\psitwos  {\ensuremath{\psi{(2S)}}\xspace}

\mathchardef\Upsilon="7107
\def\Y#1S{\ensuremath{\Upsilon{(#1S)}}\xspace}

\def\FourS {\Y4S}

\mathchardef\Deltares="7101
\mathchardef\Xi="7104
\mathchardef\Lambda="7103
\mathchardef\Sigma="7106
\mathchardef\Omega="710A

\def\Deltabar{\kern 0.25em\overline{\kern -0.25em \Deltares}{}\xspace}
\def\Lbar{\kern 0.2em\overline{\kern -0.2em\Lambda\kern 0.05em}\kern-0.05em{}\xspace}
\def\Sigbar{\kern 0.2em\overline{\kern -0.2em \Sigma}{}\xspace}
\def\Xibar{\kern 0.2em\overline{\kern -0.2em \Xi}{}\xspace}
\def\Obar{\kern 0.2em\overline{\kern -0.2em \Omega}{}\xspace}
\def\Nbar{\kern 0.2em\overline{\kern -0.2em N}{}\xspace}
\def\Xb{\kern 0.2em\overline{\kern -0.2em X}{}\xspace}

\def\btosll     {\ensuremath{b \to s \ellell}\xspace}

\def\mes        {\mbox{$m_{\rm ES}$}\xspace}

\def\DeltaE     {\mbox{$\Delta E$}\xspace}

\newcommand{\tev}{\ensuremath{\mathrm{\,Te\kern -0.1em V}}\xspace}
\newcommand{\gev}{\ensuremath{\mathrm{\,Ge\kern -0.1em V}}\xspace}
\newcommand{\mev}{\ensuremath{\mathrm{\,Me\kern -0.1em V}}\xspace}
\newcommand{\kev}{\ensuremath{\mathrm{\,ke\kern -0.1em V}}\xspace}
\newcommand{\ev}{\ensuremath{\mathrm{\,e\kern -0.1em V}}\xspace}
\newcommand{\gevc}{\ensuremath{{\mathrm{\,Ge\kern -0.1em V\!/}c}}\xspace}
\newcommand{\mevc}{\ensuremath{{\mathrm{\,Me\kern -0.1em V\!/}c}}\xspace}
\newcommand{\gevcc}{\ensuremath{{\mathrm{\,Ge\kern -0.1em V\!/}c^2}}\xspace}
\newcommand{\mevcc}{\ensuremath{{\mathrm{\,Me\kern -0.1em V\!/}c^2}}\xspace}

\def\invfb   {\ensuremath{\mbox{\,fb}^{-1}}\xspace}

\def\mus  {\ensuremath{\rm \,\mus}\xspace}

\def\mus        {\ensuremath{\,\mu{\rm s}}\xspace}    

\def\to                 {\ensuremath{\rightarrow}\xspace}

\def\pep2{PEP-II}

\newcommand{\dedx}{\ensuremath{\mathrm{d}\hspace{-0.1em}E/\mathrm{d}x}\xspace}
\newcommand{\chisq}{\ensuremath{\chi^2}\xspace}

\def\gsim{{~\raise.15em\hbox{$>$}\kern-.85em
          \lower.35em\hbox{$\sim$}~}\xspace}
\def\lsim{{~\raise.15em\hbox{$<$}\kern-.85em
          \lower.35em\hbox{$\sim$}~}\xspace}

\def\CP                {\ensuremath{C\!P}\xspace}

\xspace

\def\jetset74   {\mbox{\tt Jetset \hspace{-0.5em}7.\hspace{-0.2em}4}\xspace}

\newcommand{\gevcccc}{\ensuremath{{\mathrm{\,Ge\kern -0.1em V^2\!/}c^4}}\xspace}

\def\BToXll {\ensuremath{B\rightarrow X_{s}\, \ell^{+}\ell^{-}}\xspace}

\def\Kmaybestar {\ensuremath{K^{(*)}\xspace}}
\def\afb {\mbox{${\cal A}_{FB}$}\xspace}

\def\ctk {\ensuremath{\cos\theta_{K}}\xspace}

\def\ctl {\ensuremath{\cos\theta_{\ell}}\xspace}

\def\fl {\mbox{$F_L$}\xspace}

\def\kmaybell {\B\to\Kmaybestar\ellell\xspace}

\def\mkpi {\ensuremath{m_{\kaon\pi}}\xspace}

\def\modeeight {\ensuremath{B^0\rightarrow \Kp \pim \mumu}\xspace}
\def\modeeightshort {\ensuremath{\Kp \pim \mumu}\xspace}

\def\modeeleven {\ensuremath{B^+\rightarrow \KS \pip \epem}\xspace}
\def\modeelevenshort {\ensuremath{\KS \pip \epem}\xspace}

\def\modekstkll {\ensuremath{B^0\rightarrow K^{*0}\ellell}\xspace}

\def\modekstksll {\ensuremath{B^+\rightarrow K^{*+}\ellell}\xspace}

\def\modekstll {\ensuremath{B\rightarrow K^{*}\ellell}\xspace}

\def\modekstzll {\ensuremath{B^0\rightarrow K^{*0}\ellell}\xspace}

\def\modeseven {\ensuremath{B^+\rightarrow \KS \pip \mumu}\xspace}
\def\modesevenshort {\ensuremath{\KS \pip \mumu}\xspace}

\def\modeten {\ensuremath{B^+\rightarrow \Kp \piz \epem}\xspace}
\def\modetenshort {\ensuremath{\Kp \piz \epem}\xspace}

\def\modetwelve {\ensuremath{B^0\rightarrow \Kp \pim \epem}\xspace}
\def\modetwelveshort {\ensuremath{\Kp \pim \epem}\xspace}

\def\modekstkllshort {\ensuremath{K^{*0}(892)\ellell}\xspace}
\def\modekstksllshort {\ensuremath{K^{*+}(892)\ellell}\xspace}

\def\modekstllshort {\ensuremath{K^{*}\ellell}\xspace}

\long\def\inst#1{\par\nobreak\kern 4pt\nobreak
    {\it #1}\par\vskip 10pt plus 3pt minus 3pt}

\begin{document}

\preprint{\babar-PUB-\BABARPubYear/\BABARPubNumber}
\preprint{SLAC-PUB-\SLACPubNumber}

\begin{flushleft}
\babar-PUB-\BABARPubYear/\BABARPubNumber\\
SLAC-PUB-\SLACPubNumber\\
\vspace{\baselineskip}
\end{flushleft}

\title{
\large \bf
\boldmath
          Measurement of Angular Asymmetries in the Decays $\modekstll$
}

%
\author{J.~P.~Lees}
\author{V.~Poireau}
\author{V.~Tisserand}
\affiliation{Laboratoire d'Annecy-le-Vieux de Physique des Particules (LAPP), Universit\'e de Savoie, CNRS/IN2P3,  F-74941 Annecy-Le-Vieux, France}
\author{E.~Grauges}
\affiliation{Universitat de Barcelona, Facultat de Fisica, Departament ECM, E-08028 Barcelona, Spain }
\author{A.~Palano$^{ab}$ }
\affiliation{INFN Sezione di Bari$^{a}$; Dipartimento di Fisica, Universit\`a di Bari$^{b}$, I-70126 Bari, Italy }
\author{G.~Eigen}
\author{B.~Stugu}
\affiliation{University of Bergen, Institute of Physics, N-5007 Bergen, Norway }
\author{D.~N.~Brown}
\author{L.~T.~Kerth}
\author{Yu.~G.~Kolomensky}
\author{M.~J.~Lee}
\author{G.~Lynch}
\affiliation{Lawrence Berkeley National Laboratory and University of California, Berkeley, California 94720, USA }
% ...
\collaboration{The \babar\ Collaboration}
\noaffiliation


\vspace{\baselineskip}

\begin{abstract}
\noindent We study the lepton forward-backward asymmetry $\afb$
and the longitudinal $\Kstar$ polarization $\fl$,
as well as an observable $P_2$ derived from them,
in the rare decays $\modekstll$, where $\ellell$ is either $\epem$ or $\mumu$,
using the full sample of 471 million $\BB$ events collected at the $\FourS$ resonance with
the \babar\, detector at the \pep2\ $\epem$ collider. We
separately fit and report results for the $\modekstkllshort$ and
$\modekstksllshort$ final states, as well as their
combination $\modekstllshort$, in five disjoint dilepton mass-squared
bins. An angular analysis of $\modekstksll$ decays is presented here for
the first time.
\end{abstract}


\maketitle

\section{Introduction}
\label{sec:introduction}

\begin{figure}[h]
\begin{center}
\includegraphics[height=3.5cm]{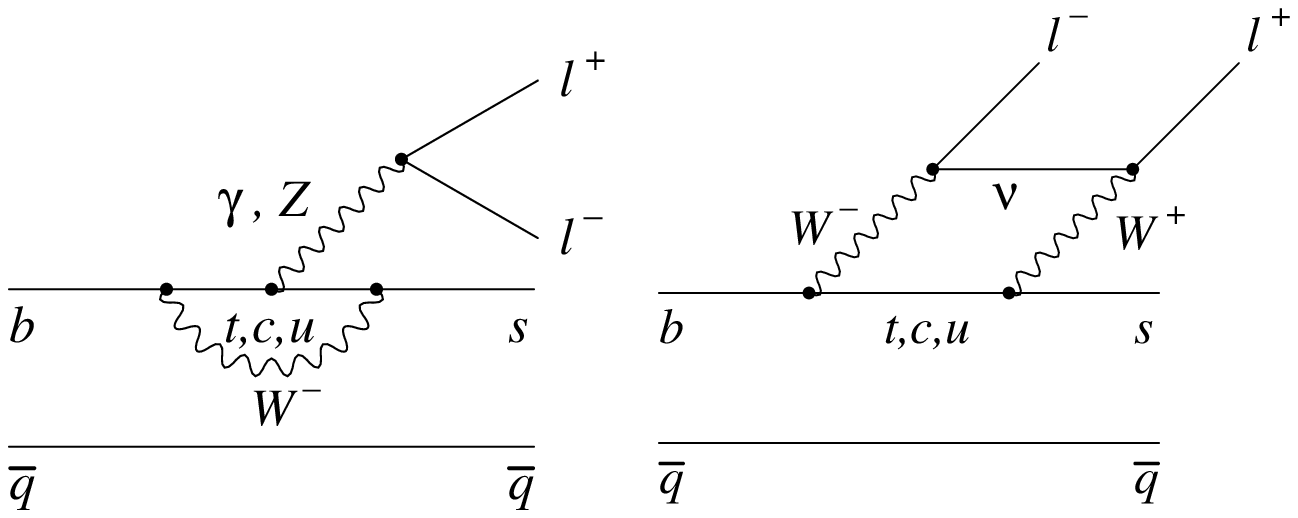}
\caption{Lowest-order SM Feynman diagrams for $\btosll$.}
\label{fig:sll_diagrams}
\end{center}
\end{figure}

\begin{figure}[]
\begin{center}
\includegraphics[]{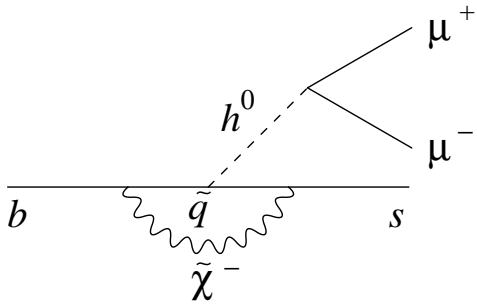}
\caption{Feynman diagram of a non-SM Higgs penguin process.}
\label{fig:sll_np}
\end{center}
\end{figure}

The decays $B\rightarrow K^{*}(892) \ellell \xspace$,
where $K^*\to K\pi$ (hereinafter, unless explicitly stated otherwise,
$K^*$ refers generically to the $K^*(892)$) and $\ellell$ is either
an $\epem$ or $\mumu$ pair,
arise from flavor-changing
neutral-current (FCNC) processes, which are forbidden at tree
level in the Standard Model (SM). The lowest-order
SM processes contributing to these decays are the
photon penguin, the $Z$ penguin and the $W^+W^-$ box
diagrams shown in Fig.~\ref{fig:sll_diagrams}. Their
amplitudes are expressed in terms of hadronic form
factors and perturbatively-calculable effective
Wilson coefficients, $C^{\rm eff}_{7}$, $C^{\rm eff}_{9}$
and $C^{\rm eff}_{10}$, which represent the
electromagnetic penguin diagram, and the vector part and
the axial-vector part of the linear combination of the $Z$
penguin and $W^+W^-$ box diagrams, respectively~\cite{Buchalla:1995vs,AFB_SMa,AFB_SMb,AFB_SMc,KrugerMatias,Altmannshofer:2008dz,Hovhannisyan}.
Non-SM physics may add new penguin and/or box diagrams,
as well as possible contributions from new scalar, pseudoscalar,
and/or tensor currents, which can contribute at the same order
as the SM diagrams, modifying the effective Wilson
coefficients from their SM
expectations~\cite{NewPhysicsA,NewPhysicsB,NewPhysicsC,NewPhysicsD,NewPhysicsE,NewPhysicsF,NewPhysicsG,NewPhysicsH,NewPhysicsI,NewPhysicsJ}.
An example of a non-SM physics loop process is shown in Fig.~\ref{fig:sll_np};
other possible processes could involve e.g., non-SM Higgs, charginos, gauginos, neutralinos
and/or squarks.
As a function of dilepton mass-squared $q^2=m_{\ellell}^2$, the
angular distributions in $\modekstll$ decays are notably sensitive
to many possible sources of new physics, with several collaborations presenting results over the past
few years~\cite{Aubert:2008bi,Wei:2009zv,Aaltonen:2011ja,Aaij:2013qta,Chatrchyan:2013cda,ATLAS:2013ola,NewPhysicsLHCb,NewPhysicsLHCbTheory}.

At any particular $q^2$ value,
the kinematic distribution of the decay products
of $\modekstll$ and the $\CP$-conjugate
$\Bbar \rightarrow \Kstarb \ellell$ process
depends on six transversity amplitudes which, neglecting
$\CP$-violating effects and terms of order $m_{\ell}^{2}$ and higher,
can be expressed as a triply differential cross-section
in three angles:
$\theta_K$, the angle between the $K$
and the $B$ directions in the $K^*$ rest frame;
$\theta_{\ell}$, the angle between the
$\ell^+(\ell^-)$ and the $B(\Bbar)$ direction in
the $\ellell$ rest frame; and $\phi$, the angle
between the $\ellell$ and $K\pi$ decay planes in
the $B$ rest frame.
From the distribution of the angle $\theta_K$ obtained
after integrating over $\phi$ and $\theta_{\ell}$,
we determine the $K^*$ longitudinal polarization
fraction $\fl$ using a fit to $\ctk$ of the form~\cite{KrugerMatias}
\begin{eqnarray}\nonumber
\frac{1}{\Gamma (q^{2})} \frac{\mathrm{d}\Gamma}{\mathrm{d}(\ctk)} &=& \frac{3}{2}{\fl (q^{2})}\cos^{2}\theta_{K} + \\
&& \frac{3}{4}(1-{\fl (q^{2})}) (1-\cos^{2}\theta_{K}) \, .
\label{eq:fl}
\end{eqnarray}
\noindent We similarly determine the
lepton forward-backward asymmetry $\afb$
from the distribution of the
angle $\theta_{\ell}$ obtained after integrating
over $\phi$ and $\theta_K$,~\cite{KrugerMatias}
\begin{eqnarray}\nonumber
\frac{1}{\Gamma (q^{2})} \frac{\mathrm{d}\Gamma}{\mathrm{d}(\ctl)} &=& \frac{3}{4}{\fl (q^{2})}(1-\cos^{2}\theta_{l}) + \\
\nonumber && \frac{3}{8}(1-{\fl (q^{2})}) (1+\cos^{2}\theta_{l}) + \\
&& {\afb (q^{2})}\cos\theta_{l} \, .
\label{eq:afb}
\end{eqnarray}
\noindent We ignore here possible contributions from non-resonant S-wave
$B \to K\pi \ellell$ events. The rate for such events has been shown
to be consistent with zero~\cite{Lees:2013nxa}, with an upper limit (68\% CL)
across the entire dilepton mass-squared range
of $<4\%$ of the $B \to \Kstar(K\pi) \ellell$ branching
fraction~\cite{Aaij:2013qta}. The presence of an S-wave component
at this level was shown to lead to a relatively small absolute bias
on the order of 0.01 for $\fl$ and $\afb$; this small bias is ignored
here given the relatively larger magnitude of our
statistical and systematic uncertainties.
Essentially no contributions from low-mass tails of the higher
$\Kstar$ resonances are expected in the $\Kstar(892)$
mass region considered here.

We ignore small $q^2$-dependent theory corrections
in the large-recoil $q^2 \lesssim 2 \gevcccc$ region given
the current experimental uncertainties on the angular observables, which are relatively
large compared to these small corrections in the underlying
SM theory expectations~\cite{Altmannshofer:2008dz}.
We determine $\fl$ and $\afb$ in the five disjoint bins of
$q^2$ defined in Table~\ref{tab:sbins}.
We also present results in a
$q^2$ range $1.0 < q^{2}_{0} < 6.0 \gevcccc$, the perturbative window away
from the $q^{2} \to 0$ photon pole and the $\ccbar$ resonances
at higher $q^{2}$, where theory uncertainties are considered to be under good control.
An angular analysis of the decays $\modekstksll$ is presented here for the first time.
We additionally present results for an observable
derived from $\fl$ and $\afb$, $P_2 = (-2/3) * \afb / (1-\fl)$,
with less theory uncertainty, and hence greater sensitivity
to non-SM contributions, than either $\fl$ or $\afb$
alone~\cite{NewPhysicsP2,dc}.

\begin{table}[b!]
\centering
\caption[$q^2$ bins]{\label{tab:sbins}
Definition of the $q^2$ bins used in the analysis.
The nominal $B$ and $\Kstar$ invariant masses~\cite{Agashe:2014kda} are
given by $m_B$ and $m_{\Kstar}$, respectively.}
\begin{tabular}{ccccc}
\\
\hline\hline
\\
$q^{2}$ bin & & $q^2$ min (\gevcccc) & & $q^2$ max (\gevcccc) \\
\noalign{\vskip 1mm}
\hline
\noalign{\vskip 1mm}
$q^{2}_1$   & & 0.10                 & & 2.00                        \\
$q^{2}_2$   & & 2.00                 & & 4.30                        \\
$q^{2}_3$   & & 4.30                 & & 8.12                        \\
$q^{2}_4$   & & 10.11                & & 12.89                       \\
$q^{2}_5$   & & 14.21                & & $(m_B-m_{\Kstar})^{2}$ \\
\noalign{\vskip 1mm}
\hline
\noalign{\vskip 1mm}
$q^{2}_0$   & & 1.00                 & & 6.00                        \\
\noalign{\vskip 1mm}
\hline
\hline
\end{tabular}
\end{table}

\section{Event Selection}
\label{sec:dataset}

We use a data sample of $\sim 471$ million $\BB$ pairs,
corresponding to $424.2 \pm 1.8 \invfb$~\cite{babarlumi},
collected at the $\FourS$ resonance with the \babar\, detector~\cite{BaBarDetector}
at the PEP-II asymmetric-energy $\epem$ collider at the SLAC
National Accelerator Laboratory. Charged particle tracking is
provided by a five-layer silicon vertex tracker and a 40-layer
drift chamber in a 1.5 T solenoidal magnetic field. We identify
electrons and photons with a CsI(Tl) electromagnetic calorimeter, and muons
using an instrumented magnetic flux return. We identify charged
kaons using a detector of internally reflected Cherenkov light,
as well as $\dedx$ information from the drift chamber. Charged
tracks other than identified $e$, $\mu$ and $K$ candidates are
treated as pions.

We reconstruct $\modekstll$ signal events in the following
final states (charge conjugation is implied throughout unless
explicitly noted):
\begin{itemize}
\item $\Bp\to \Kstarp(\to \KS\pip)\mumu$;
\item $\Bz\to \Kstarz(\to \Kp\pim)\mumu$;
\item $\Bp\to \Kstarp(\to \Kp\piz)\epem$;
\item $\Bp\to \Kstarp(\to \KS\pip)\epem$;
\item $\Bz\to \Kstarz(\to \Kp\pim)\epem$.
\end{itemize}
We do not include the decays
$\Bp\to \Kstarp(\to \Kp\piz)\mumu$ and
$\Bz\to \Kstarz(\to \KS\piz)\ellell$
in our analysis. The expected signal-to-background ratio for
these final states relative to the five chosen signal
modes listed above is very poor, with ensembles of pseudo-experiments showing that
inclusion of these extra modes would yield no additional sensitivity.

We require $\Kstar$ candidates to have an invariant
mass $0.72 < m(K\pi) < 1.10 \gevcc$. Electron and muon candidates
are required to have momenta $p > 0.3 \gevc$ in the laboratory frame.
The muon and electron misidentification rates determined from high-purity data
control samples are, respectively, $\sim 2\%$ and $\lesssim 0.1\%$~\cite{BaBarDetector}, and
backgrounds from particle misidentification are thus significant for $B \to \Kstar \mumu$ candidates only.
We combine up to three photons with an electron candidate when the
photons are consistent with bremsstrahlung from the electron.
We do not use electrons that are associated with
photon conversions to low-mass $\epem$ pairs. We reconstruct $\KS$
candidates in the $\pip\pim$ final state, requiring an invariant
mass consistent with the nominal $K^0$ mass, and a flight distance
from the $\epem$ interaction point that is more than three times the flight
distance uncertainty. Neutral pion candidates are formed from two photons with
$E_{\gamma} > 50 \mev$, and an invariant mass between $115$ and
$155 \mevcc$. In each final state, we utilize the kinematic variables
$\mes=\sqrt{E^2_{\rm CM}/4 -p^{*2}_B}$ and $\DeltaE = E_B^* - E_{\rm CM}/2$,
where $p^*_B$ and $E_B^*$ are the $B$ momentum and energy in the $\Upsilon(4S)$
center-of-mass (CM) frame, and $E_{\rm CM}$ is the total CM energy.
We reject events with $\mes < 5.2 \gevcc$.

To characterize backgrounds from hadrons misidentified as muons,
we study $\Kstar h^\pm \mu^\mp$ candidates,
where $h$ is a charged track with no particle identification
requirement applied. We additionally use a $K^* e^\pm \mu^\mp$ sample,
where no signal is expected because of lepton-flavor conservation,
to model the combinatorial background from two random leptons.
For both $\epem$ and $\mumu$ modes, we veto the $\jpsi$ $(2.85 < m_{\ellell} < 3.18 \gevcc)$
and $\psitwos$ $(3.59 < m_{\ellell} < 3.77 \gevcc)$ mass regions.
These vetoed events provide high-statistics control samples
of decays to final states identical to the signal modes here
that we use to validate our fitting procedures.

Random combinations of leptons from semileptonic $B$ and $D$ decays
are the predominant source of backgrounds.
These combinatorial backgrounds occur in both $\BB$ events
(``$\BB$ backgrounds'') and $\epem \to \qqbar$ continuum events
(``$\qqbar$ backgrounds'', where $q=u,d,s,c$), and are suppressed using eight bagged decision trees (BDTs)~\cite{bdts}
trained for suppression of:
\begin{itemize}
\item $\BB$ backgrounds in $\epem$ modes at low $q^2$;
\item $\BB$ backgrounds in $\epem$ modes at high $q^2$;
\item $\BB$ backgrounds in $\mumu$ modes at low $q^2$;
\item $\BB$ backgrounds in $\mumu$ modes at high $q^2$;
\item $\qqbar$ backgrounds in $\epem$ modes at low $q^2$;
\item $\qqbar$ backgrounds in $\epem$ modes at high $q^2$;
\item $\qqbar$ backgrounds in $\mumu$ modes at low $q^2$;
\item $\qqbar$ backgrounds in $\mumu$ modes at high $q^2$,
\end{itemize}
\noindent where low (high) $q^2$ is defined as the mass-squared region below (above)
the vetoed $\jpsi$ region. In order to treat the $K^* e^\pm \mu^\mp$ control sample
equivalently to the $\epem$ and $\mumu$ datasets, we similarly train four BDTs for
$\BB$ and $\qqbar$ background suppression in the low and high $q^2$ regions, using a
high-statistics sample of simulated $B \to \Kstar e^\pm \mu^\mp$ events. The
$\mumu$ BDTs are used to characterize the $\Kstar h^\pm \mu^\mp$ dataset.

\begin{figure*}
\centering
\subfigure[~$\epem$ BDT output for $\BB$ background suppression in $\modeeleven$.]{
\label{fig3a}
\includegraphics[width=.45\textwidth]{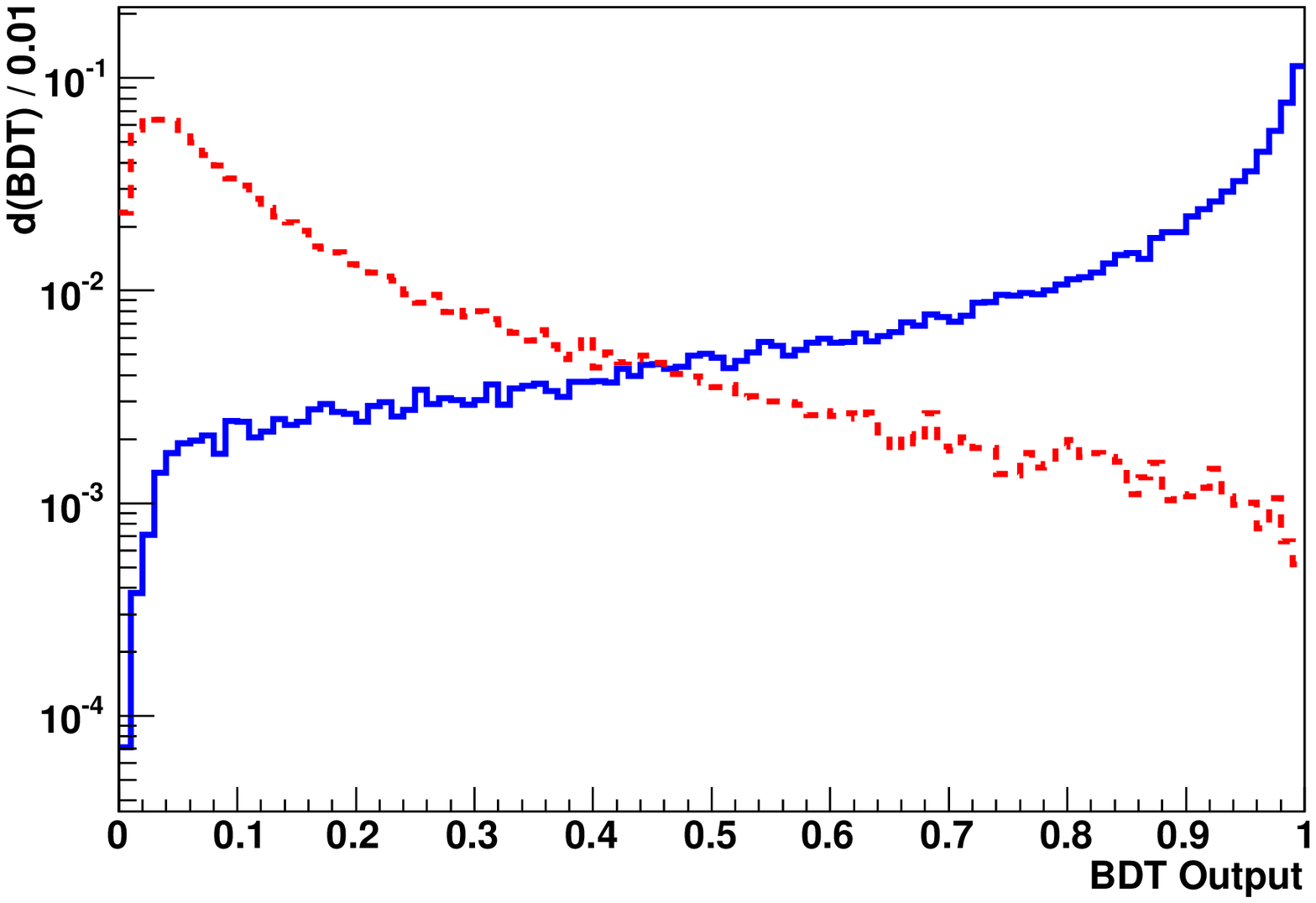}}
\subfigure[~$\epem$ BDT output for $\qqbar$ background suppression in $\modeeleven$.]{
\label{fig3b}
\includegraphics[width=.45\textwidth]{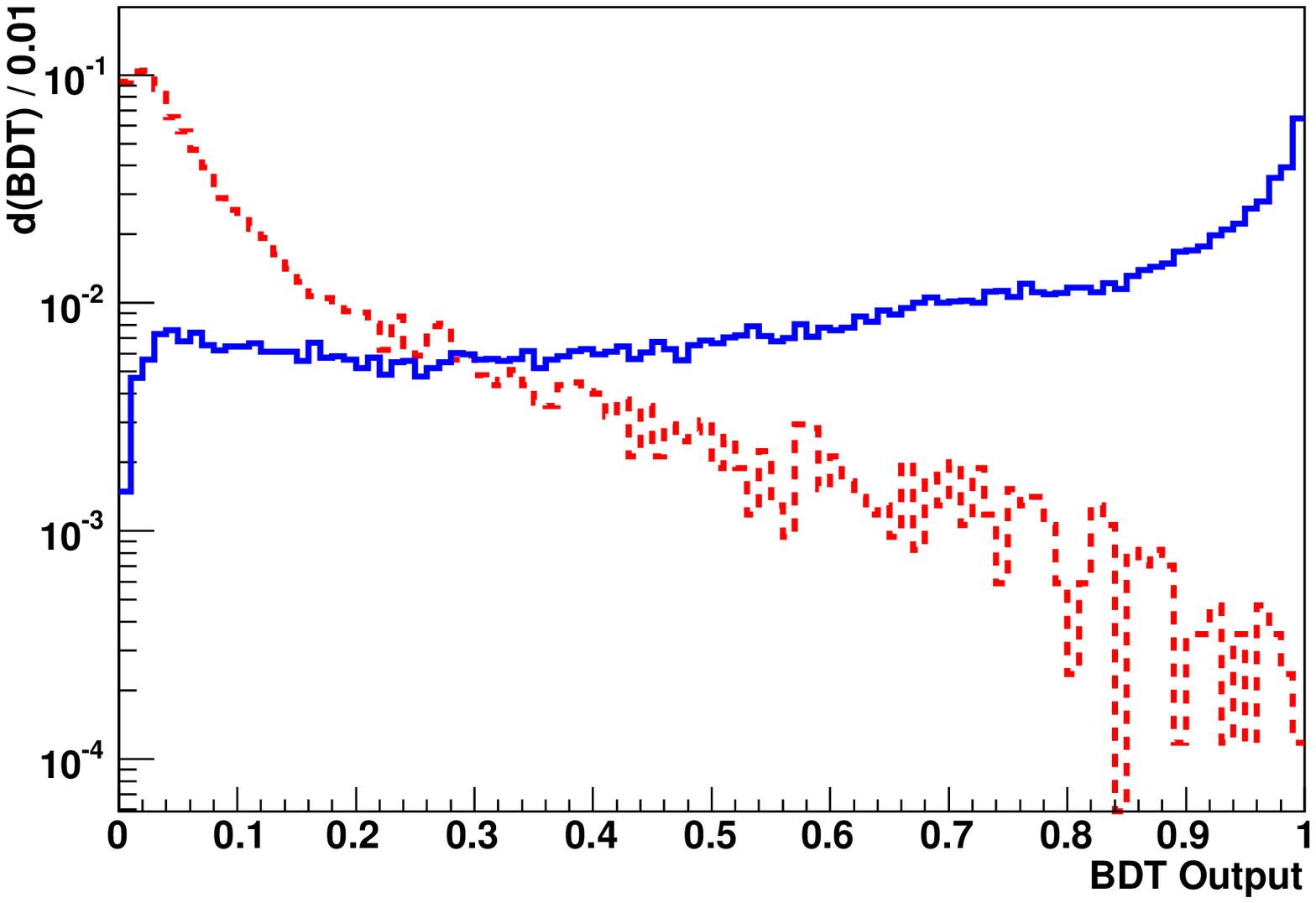}}
\subfigure[~$\mumu$ BDT output for $\BB$ background suppression in $\modeseven$.]{
\label{fig3c}
\includegraphics[width=.45\textwidth]{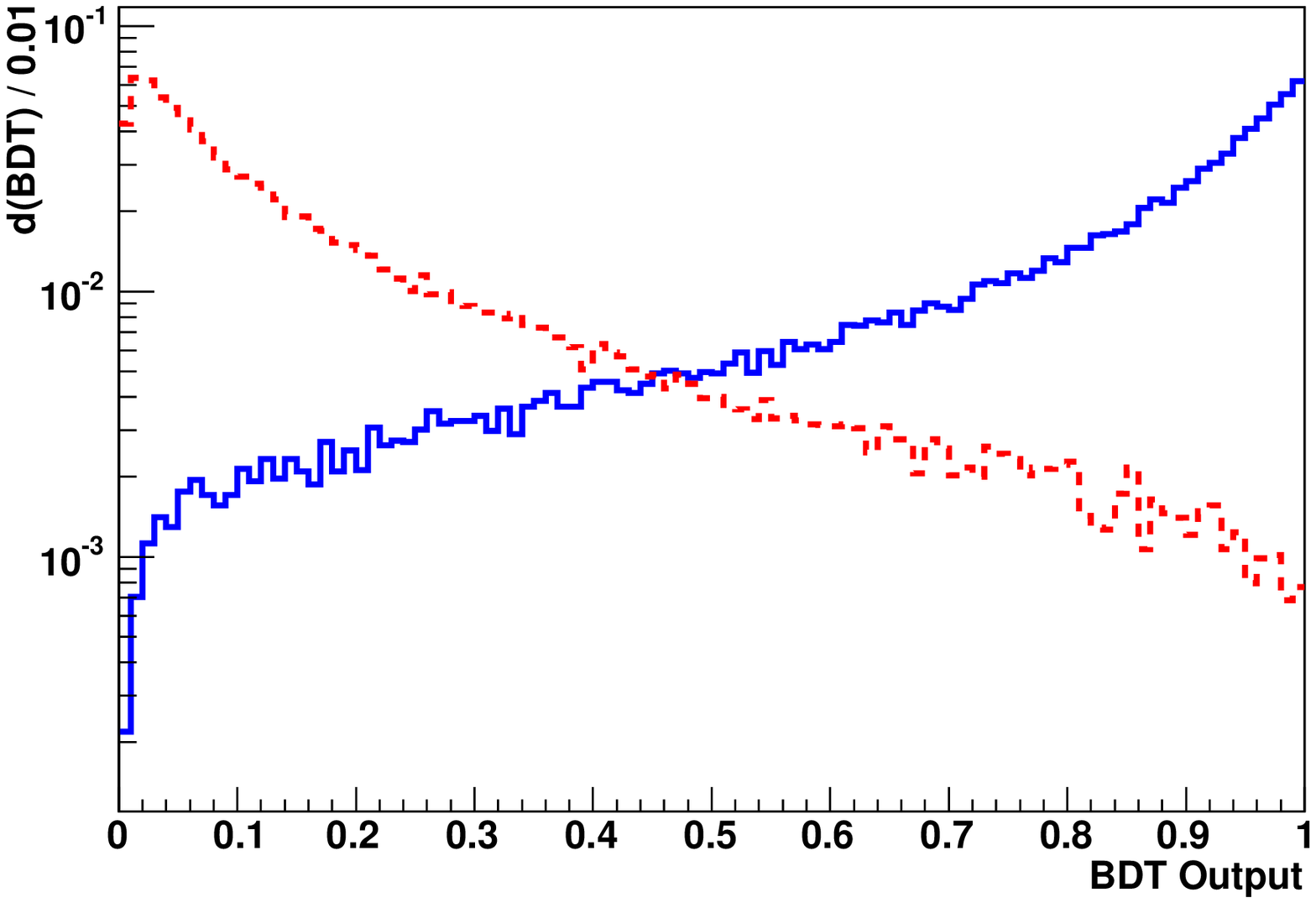}}
\subfigure[~$\mumu$ BDT output for $\qqbar$ background suppression in $\modeseven$.]{
\label{fig3d}
\includegraphics[width=.45\textwidth]{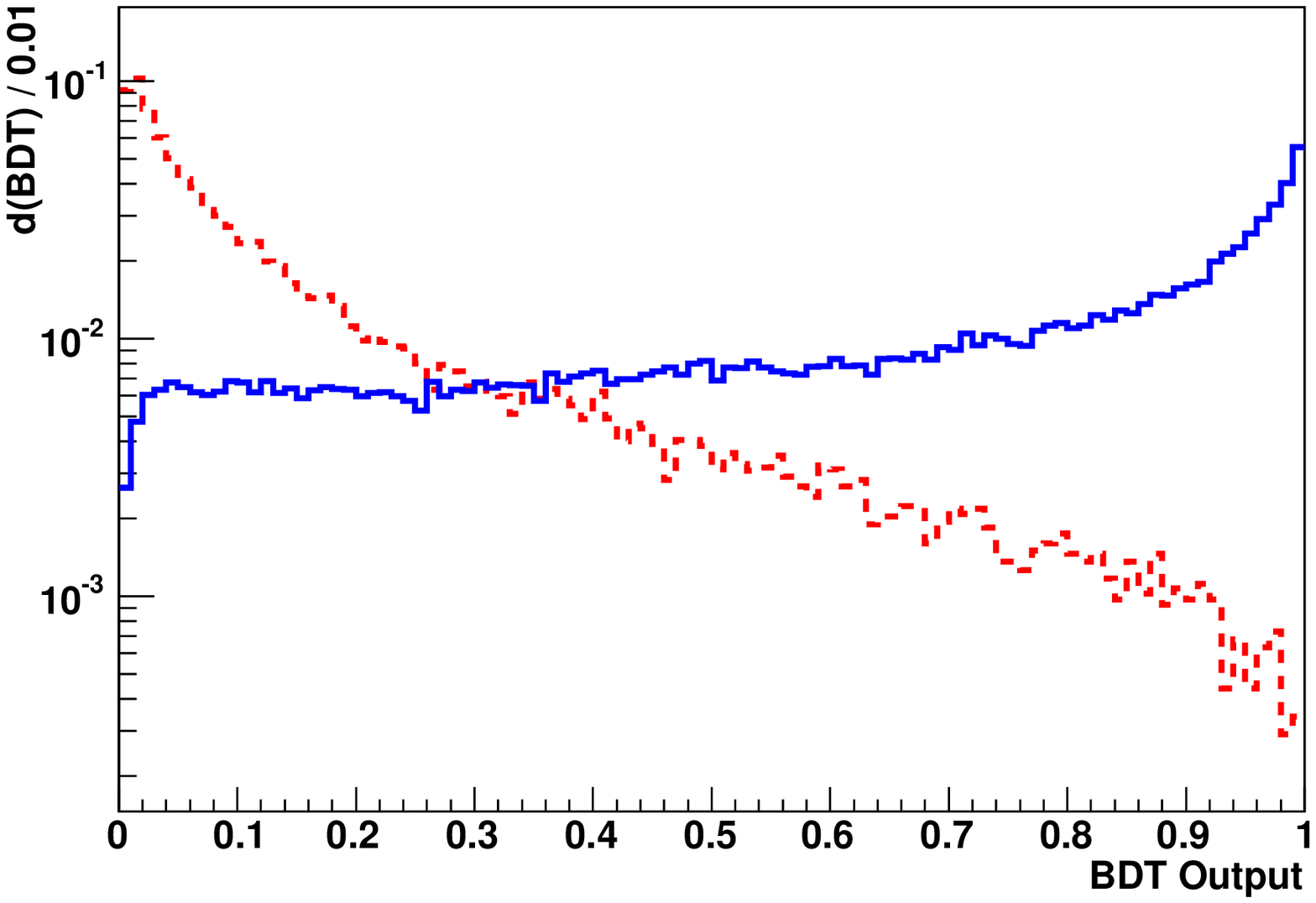}}
\caption{BDT outputs normalized to unit area for
simulated signal (solid blue line) and background (red dashed line) $q^{2}_1$ events.}
\label{fig3}
\end{figure*}

Each of the above BDTs uses a subset of the following observables
as its input parameters:
\begin{itemize}
\item the $\B$ candidate $\DeltaE$;
\item the ratio of Fox-Wolfram moments $R_2$~\cite{Foxwolfram} and the ratio of
the second-to-zeroth angular moments of the energy flow $L_2/L_0$~\cite{LegMom},
both of which are event shape parameters calculated using charged and neutral particles in the CM frame;
\item the mass and $\DeltaE$ of the other $B$ meson in the event
computed in the laboratory
frame by summing the momenta and energies of all charged particles
and photons that are not used to reconstruct the signal candidate;
\item the magnitude of the total transverse momentum of the event;
\item the $\chisq$ probability of the vertex fitted from all the B candidate tracks;
\item the cosines of four angles, all defined in the CM frame:
the angle between the $B$ candidate momentum and the beam axis,
the angle between the event thrust axis and the beam axis,
the angle between the thrust axis of the rest of the event and the beam axis, and
the angle between the event thrust axis and the thrust axis of the rest of the event.
The thrust $T$ of an event comprised of $N$ particles, or analogously for a subset of particles
in an event, is defined as~\cite{Brandt:1978zm}
\begin{equation}\nonumber
T = \frac{\sum\limits_{i=1}^{N} |\vec{p}_{i} \cdot \hat{t}|}{\sum\limits_{i=1}^{N} |\vec{p}_{i}|} \, ,
\label{eq:thrust}
\end{equation}
\noindent where the thrust axis $\hat{t}$ maximizes the magnitude of the thrust $T$, up to
a two-fold ambiguity in direction (forward and backward are equivalent).
\end{itemize}
As an example, Fig.~\ref{fig3} shows histograms of BDT output normalized to unit area
for simulated $\modeelevenshort$ and $\modesevenshort$ signal and combinatorial background
events in the $q^{2}_1$ bin. The BDT outputs for the other
final states and $q^2$ bins demonstrate similar discriminating power.

Backgrounds from $B \to D(\to \Kmaybestar \pi) \pi$ hadronic decays occur if two
hadrons are misidentified as leptons, which happens at a non-negligible rate only
in dimuon final states.
These events are vetoed by requiring the invariant mass of the $\Kstar\pi$
system to be outside the range $1.84-1.90 \gevcc$ after assigning the pion
mass hypothesis to the muon candidates. Residual muon misidentification
backgrounds remaining after this selection are characterized using the
$\Kstar h^\pm \mu^\mp$ dataset.

For the last steps in the event selection, we adopt
(a) the $\DeltaE$ regions used in our recent
related analyses of rates and rate asymmetries in
exclusive $\kmaybell$ and inclusive $\BToXll$ decays~\cite{Lees:2012tva,Lees:2013nxa},
$-0.1 (-0.05) < \DeltaE < 0.05 \gev$ for $\epem$ ($\mumu$) modes;
and (b) the $\qqbar$ BDT $> 0.4$ selection used in the
inclusive $\BToXll$ analysis~\cite{Lees:2013nxa}. After
all other selection criteria have been imposed, this
$\qqbar$ BDT selection removes most $\qqbar$ background
events with only trivial decreases in signal efficiencies.

At the conclusion of the event selection process, some events
have multiple reconstructed $B$ candidates which typically differ
by one charged or neutral pion in the hadronic system.
The signal candidate multiplicity averaged across final states and
$q^2$ bins is $\sim 1.4$ ($\sim 1.1$) candidates per event
in dielectron (dimuon) modes.
In events with multiple signal candidates,
the candidate with the $\DeltaE$ value closest to zero is selected.

\section{Angular Observables Extraction Method}

\subsection{General Strategy}

We extract the angular observables $\fl$ and $\afb$ from the data using a series of
likelihood (LH) fits which proceed in several steps:

\begin{itemize}
\item In each $q^2$ bin, for each of the five signal modes separately and using
the full $\mes > 5.2 \gevcc$ dataset,
an initial unbinned maximum LH fit of $\mes$, $m(K\pi)$ and a likelihood ratio
($L_R$, defined below in Eq.~\ref{eq:lhratio}) that discriminates against random combinatorial
$\BB$ backgrounds is performed. After this first fit, all normalizations and
the $\mes$-dependent, $m(K\pi)$-dependent and $L_R$-dependent probability density function (pdf) shapes
are fixed.

\item Second, in each $q^2$ bin and for each of the five signal modes
separately, $\mes$, $m(K\pi)$ and $L_R$ pdfs and normalizations
are defined for $\mes > 5.27 \gevcc$ events (the ``$\mes$ angular fit region'')
using the results of the prior three-dimensional fits.
Only $\mes$ angular fit region events and pdfs are subsequently used in
the fits for $\fl$ and $\afb$.

\item Next, $\ctk$ is added as a fourth dimension
to the likelihood function,
in addition to $\mes$, $m(K\pi)$ and $L_R$, and
four-dimensional likelihoods with $\fl$ as the only free
parameter are defined for $\mes$ angular fit region events.
As above, each $q^2$ bin and each of the five signal
modes has its own separate 4-d LH function. However, a common value
of $\fl$ is shared among all of the 4-d LH functions in any given $q^{2}$ bin. Thus, by combining
LH functions from multiple final states,
it becomes possible to extract $\fl$ and $\afb$ for arbitrary combinations of the five
final states here. In particular, we quote results using
three different sets of our five signal modes:
\begin{itemize}
\item $\modekstksll$, comprised of
  \subitem $\Bp\to \Kstarp(\to \KS\pip)\mumu$,
  \subitem $\Bp\to \Kstarp(\to \Kp\piz)\epem$,
  \subitem $\Bp\to \Kstarp(\to \KS\pip)\epem$,
\item $\modekstkll$, comprised of
  \subitem $\Bz\to \Kstarz(\to \Kp\pim)\mumu$,
  \subitem $\Bz\to \Kstarz(\to \Kp\pim)\epem$.
\item $\modekstll$, comprised of
  \subitem $\Bp\to \Kstarp(\to \KS\pip)\mumu$,
  \subitem $\Bz\to \Kstarz(\to \Kp\pim)\mumu$,
  \subitem $\Bp\to \Kstarp(\to \Kp\piz)\epem$,
  \subitem $\Bp\to \Kstarp(\to \KS\pip)\epem$,
  \subitem $\Bz\to \Kstarz(\to \Kp\pim)\epem$.
\end{itemize}
\item In the final step, we use the fitted value of $F_L$ from
the previous fit step as input to a similar 4-d fit for $\afb$,
in which $\ctl$ replaces $\ctk$ as the fourth dimension
in the LH function, in addition to $\mes$, $m(K\pi)$ and $L_R$.
\end{itemize}

As mentioned above, we define a likelihood ratio $L_R$ as
the third dimension in the initial fit,
\begin{equation}
\label{eq:lhratio}
L_R \equiv \frac{{\cal P}_{{\rm sig}}                       }
                     {{\cal P}_{{\rm sig}} + {\cal P}_{{\rm bkg}}} ,
\end{equation}
\noindent where ${\cal P}_{{\rm sig}}$ and ${\cal P}_{{\rm bkg}}$ are probabilities
calculated from the $\BB$ BDT output for signal and $\BB$ backgrounds,
respectively. ${\cal P}_{{\rm sig}}$ and ${\cal P}_{{\rm bkg}}$
are modeled using several different functional forms depending
on $q^2$ bin and final state.
After the multiple candidate selection described at the conclusion
of the preceding section and before fitting a dataset, a final
requirement of $L_R>0.6$ is made. This drastically reduces
the number of background events at the cost of a relatively
small loss, dependent on final state and $q^2$ bin, in signal efficiency.
Table~\ref{tab:finaleff} shows final signal efficiencies in the $\mes$ angular fit region
for each final state and $q^2$ bin.

\begin{table}
\centering
\caption[Signal efficiencies by mode and $q^2$ bin.]
{\label{tab:finaleff}
Final signal efficiencies in the $\mes$ angular fit region
by mode and $q^2$ bin.}
\begin{tabular}{l @{\extracolsep{1em}} c @{\extracolsep{1em}} c @{\extracolsep{1em}} c @{\extracolsep{1em}} c @{\extracolsep{1em}} c @{\extracolsep{1em}} c}
\\ \hline \hline \\
Mode & $q^2_0$ & $q^2_1$ & $q^2_2$ & $q^2_3$ & $q^2_4$ & $q^2_5$ \\ \hline
\noalign{\vskip 1mm}
$\modesevenshort$  & 0.143 & 0.130 & 0.146 & 0.145 & 0.143 & 0.108 \\
$\modeeightshort$  & 0.184 & 0.152 & 0.185 & 0.195 & 0.194 & 0.157 \\
$\modetenshort$    & 0.121 & 0.105 & 0.124 & 0.121 & 0.110 & 0.075 \\
$\modeelevenshort$ & 0.182 & 0.160 & 0.185 & 0.174 & 0.151 & 0.109 \\
$\modetwelveshort$ & 0.230 & 0.195 & 0.233 & 0.229 & 0.209 & 0.151 \\ \hline \hline
\end{tabular}
\end{table}

The initial 3-d fit is an unbinned maximum likelihood fit with
minimization performed by MINUIT~\cite{James:1975dr}.
Each angular result is subsequently determined by direct construction and
examination of the negative log-likelihood (NLL) curves
resulting from a scan across the entire $\fl$ or $\afb$
parameter space, including unphysical regions which provide
a statistically consistent description of the data.

\subsection{Event Classes}

We characterize
$\mes$, $m(K\pi)$, $L_R$, $\ctk$ and $\ctl$
probability density functions in our
likelihood fit model for several classes of events:
\begin{itemize}
\item correctly reconstructed (``true'') signal events;
\item misreconstructed (``crossfeed'') signal events, from both
the five signal modes as well as from other $\btosll$ decays;
\item random combinatorial backgrounds;
\item backgrounds from $\jpsi$ and $\psitwos$ decays which
escape the dilepton mass veto windows;
\item for the $\mumu$ modes only, backgrounds from hadronic
decays in which there is muon misidentification of hadrons
(this background is negligible in $\epem$ final states due to
the much smaller, relative to muons, electron misidentification probability).
\end{itemize}

\subsubsection{True and Crossfeed Signal Events}

True signal events have all final state
daughter particles correctly reconstructed.
The true signal normalization for each final state
in each $q^2$ bin is a free parameter
in the initial 3-d fits.
For each final state, the $\mes$ signal pdf is parameterized
as a Gaussian with a mean and width fixed to values obtained
from a fit to the vetoed $\jpsi$ data events in the same final state.
Similarly, for the resonant $\Kstar$ lineshape in each final state, the signal
$m(K\pi)$ pdf uses a relativistic Breit-Wigner (BW) with width and
pole mass fixed from the vetoed $\jpsi$ data events in the same final state.
True signal $L_R$ pdfs for each final state in each $q^2$ bin
are derived from simulated signal events.
There is good agreement between the $L_R$ shapes derived from simulated events
and the $L_R$ shapes observed in the charmonium control sample data.

Equations~\ref{eq:fl} and~\ref{eq:afb}, showing the dependence of
$\fl$ and $\afb$ on $\ctk$ and $\ctl$ respectively, are purely
theoretical expressions which must be modified to take into account
the experimental acceptance. We characterize the angular acceptance
using simulated signal events to obtain parameterizations of the
$\ctk$ and $\ctl$ efficiency for each final state in each $q^2$ bin.

Signal crossfeed typically occurs when a low-energy $\pi^{\pm}$ or $\piz$ is
swapped, added or removed from the set of daughter particles used to reconstruct
an otherwise correctly reconstructed signal candidate. There can be
self-crossfeed within one signal mode, feed-across between two different signal
modes with the same final state particle multiplicity, or (up) down crossfeed
from (lower) higher multiplicity $s\ellell$ modes. Simulated signal events are used
to model these types of decays, with normalization relative to the fitted true signal yield.
Averaged over the five signal modes and disjoint $q^2$ bins $q^2_1 - q^2_5$,
the fraction of crossfeed events relative to correctly reconstructed signal decays
is $\sim 0.4$ for events in the $\mes > 5.27 \gevcc$ angular fit region. Generator-level
variations in the production of cross-feed events are considered as part
of the study of systematic uncertainties related to the modeling of
signal decays.

\subsubsection{Combinatorial Backgrounds}

The largest source of background is from semileptonic $B$ and $D$ decays, where
leptons from two such decays and a $\Kstar$ candidate combine to form
a $B$ candidate. The $\mes$ pdf for the combinatorial
background is modeled with a kinematic threshold function~\cite{argus}
whose single shape parameter is a free parameter in
the fits. Events in the lepton-flavor violating (LFV) modes $K^* e^\pm \mu^\mp$, which
are forbidden in the SM and for which stringent experimental limits
exist~\cite{Agashe:2014kda}, are reconstructed and selected
analogously to the final event selection in order to characterize the
combinatorial background $m(K\pi)$ and $L_R$ pdfs.
We obtain the angular pdfs for the combinatorial backgrounds in
the $\mes$ angular fit region using events in the $\mes$ sideband
region $5.2 < \mes < 5.27 \gevcc$. The LFV events additionally provide an alternative model
for the combinatorial angular pdfs, which is used in the characterization of
systematic uncertainties in the angular fits.

\subsubsection{Charmonium and Other Physics Backgrounds}

Some misreconstructed charmonium events escape the charmonium vetoes and appear in our $q^2$ bins.
This typically occurs through bremsstrahlung by electrons, followed by
incorrect recovery of the missing energy.
The pdfs for this residual charmonium background are modeled
using simulated charmonium signal events.

In order to use the vetoed charmonium events as a data control sample, we construct a set of pdfs
equivalent to those used in the $\modekstll$ angular fits but which are appropriate for $\jpsi$
and $\psitwos$ events inside, rather than outside, their respective vetoed mass windows. The BDTs in the low
(high) $q^2$ bin are used to calculate $L_R$ for events within the $\jpsi$ ($\psitwos$) mass window.

Gamma conversions from $\B\to\Kstar\gamma$ events and Dalitz decays $(\piz,\eta)
\to \epem \gamma$ of hadronic $B$ decay daughters give rise to small
backgrounds in $q^2_1$. However, since less than a single event from
these sources is expected in the final angular fits, we do not include
them in our fit model.

\subsubsection{Muon Misidentification Backgrounds}

In dimuon modes only, some events pass the final selection but have
misidentified hadron(s) taking the place of one or both muon candidates.
To model these events, we follow a procedure similar to that
described in Ref.~\cite{Aubert:2006vb} by selecting a sample of $\Kstar\mu^{\pm} h^{\mp}$ events
requiring that the $\mu^{\pm}$ candidate be identified as a muon and the
$h^{\mp}$ candidate fail identification as an electron.
Using weights obtained from data control samples where a charged
particle's species can be identified with high precision and
accuracy without using particle identification information,
the $\Kstar\mu^{\pm} h^{\mp}$ dataset is weighted event-by-event to characterize
expected contributions in our fits due to the presence of misidentified muon candidates.
The pdfs for these events are implemented
as a sum of weighted histograms, with
normalizations obtained by construction
directly from the weighted control sample data.

\subsection{Initial $\mes$, $m(K\pi)$ and $L_R$ Fit}

\begin{figure*}
\centering
\subfigure[~$\mes$: \modetwelve.]{\label{fig:mesee}\includegraphics[width=.32\textwidth]{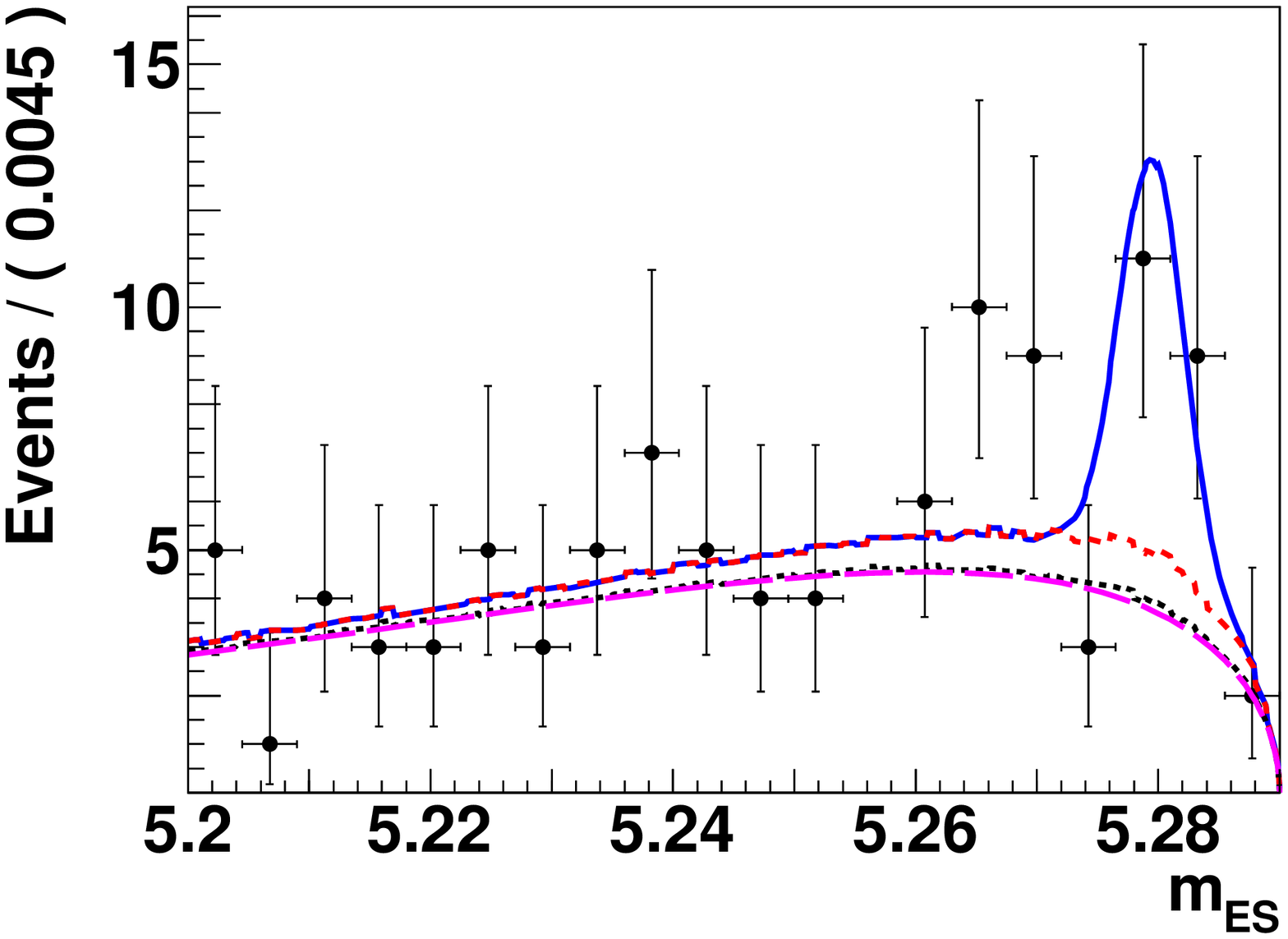}}
\subfigure[~$\mkpi$: \modetwelve.]{\label{fig:mhadee}\includegraphics[width=.32\textwidth]{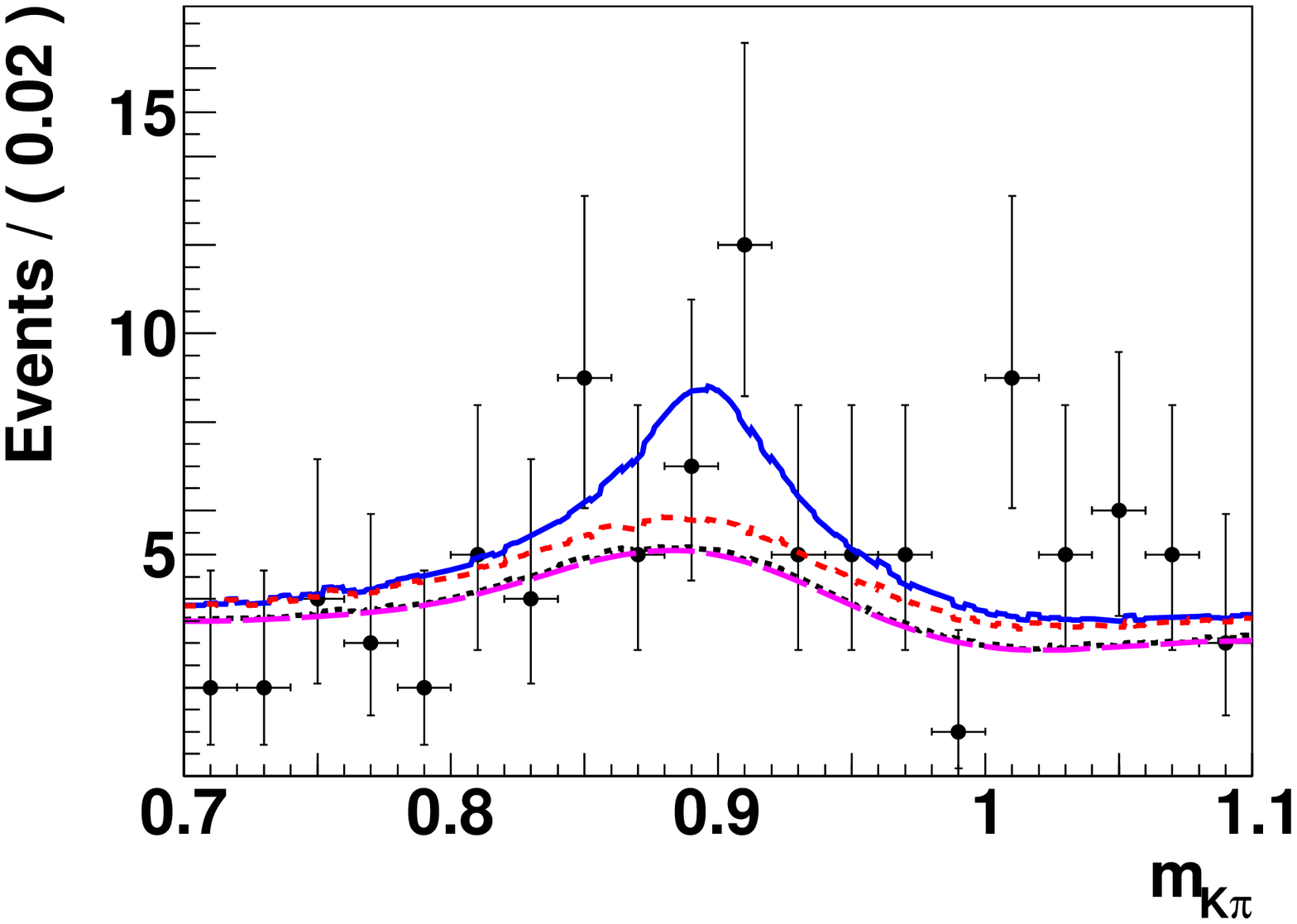}}
\subfigure[~$L_R$: \modetwelve.]{\label{fig:lhree}\includegraphics[width=.32\textwidth]{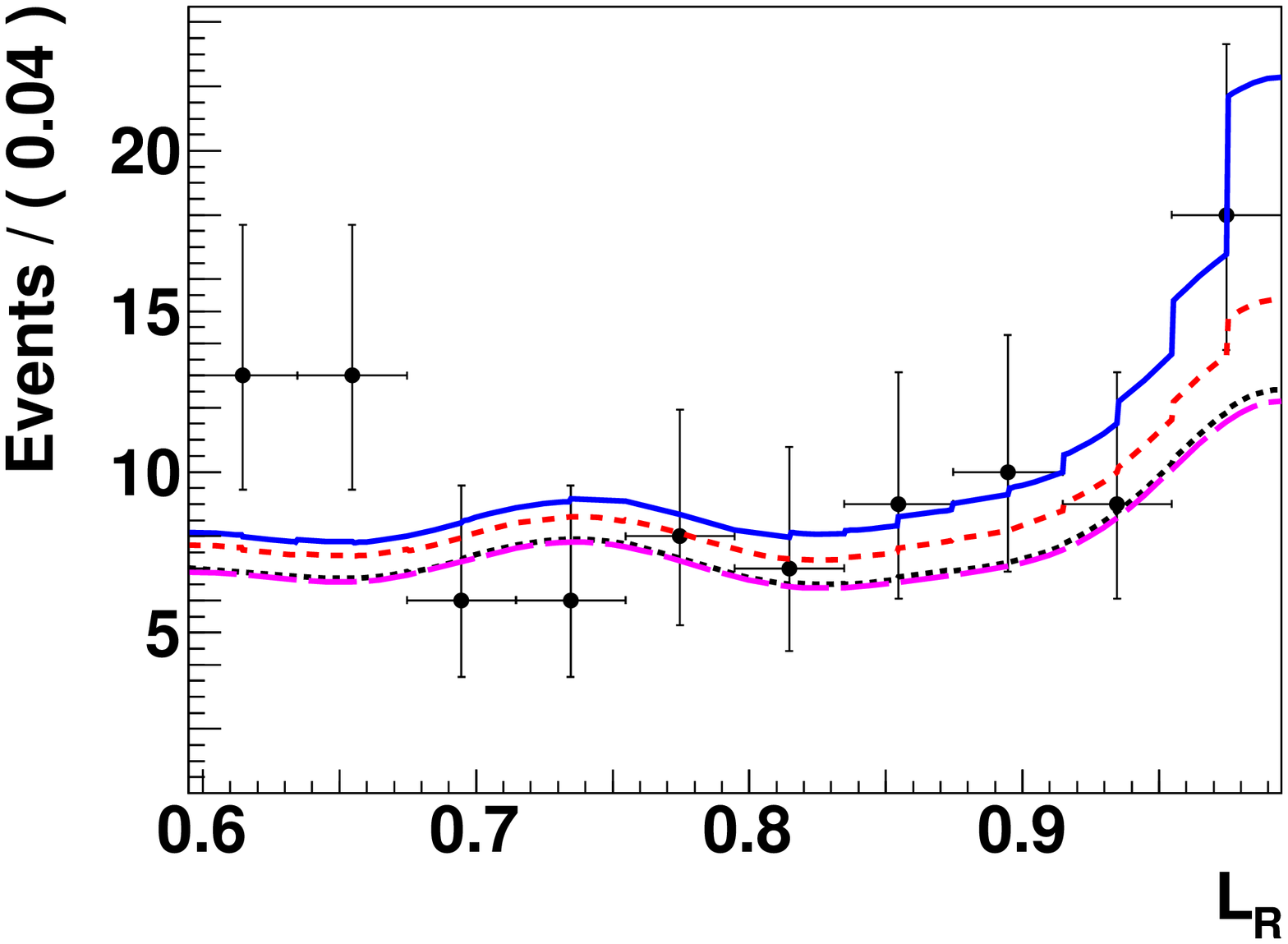}}
\subfigure[~$\mes$: \modeeight.]{\label{fig:mesmm}\includegraphics[width=.32\textwidth]{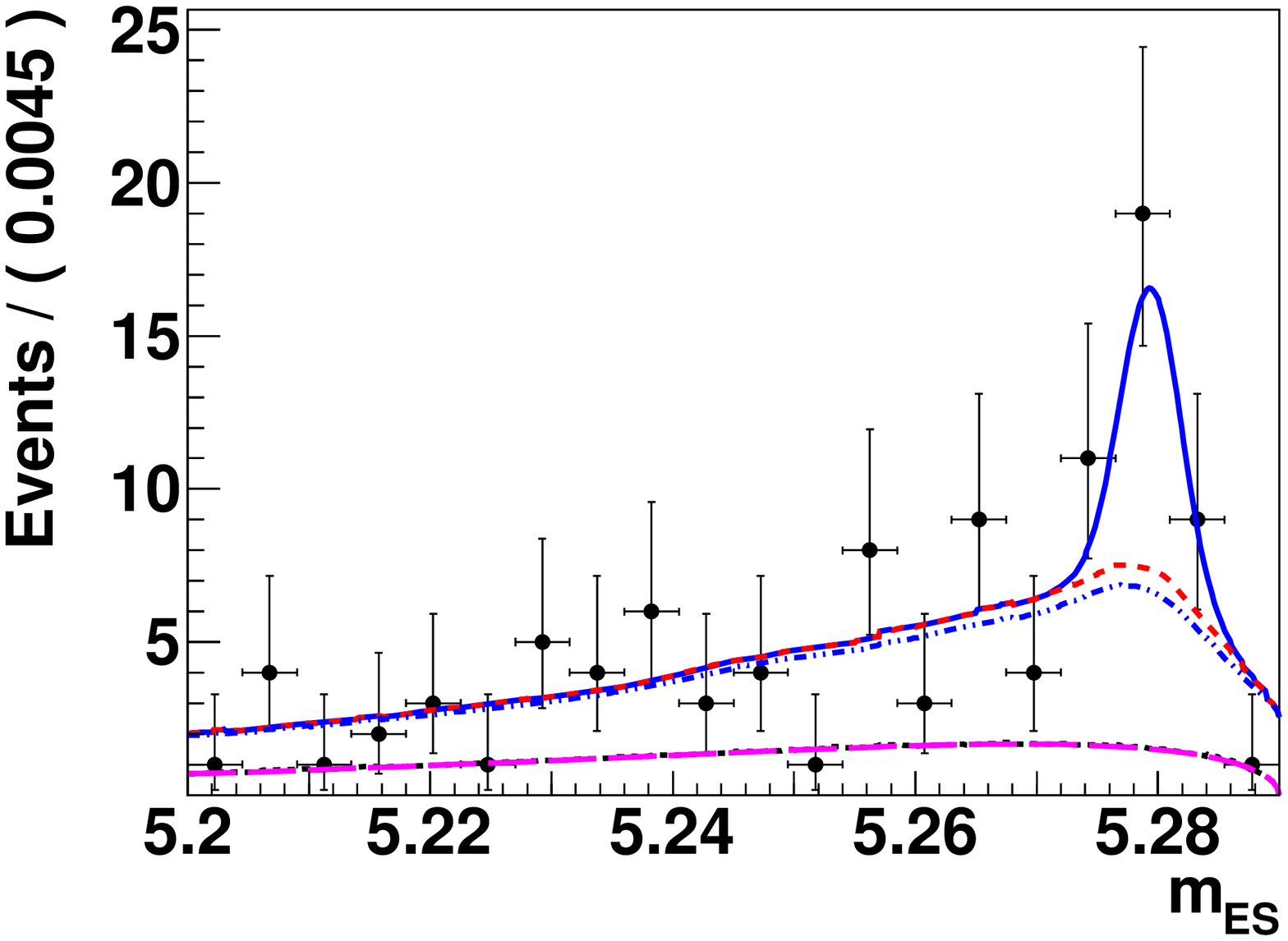}}
\subfigure[~$\mkpi$: \modeeight.]{\label{fig:mhadmm}\includegraphics[width=.32\textwidth]{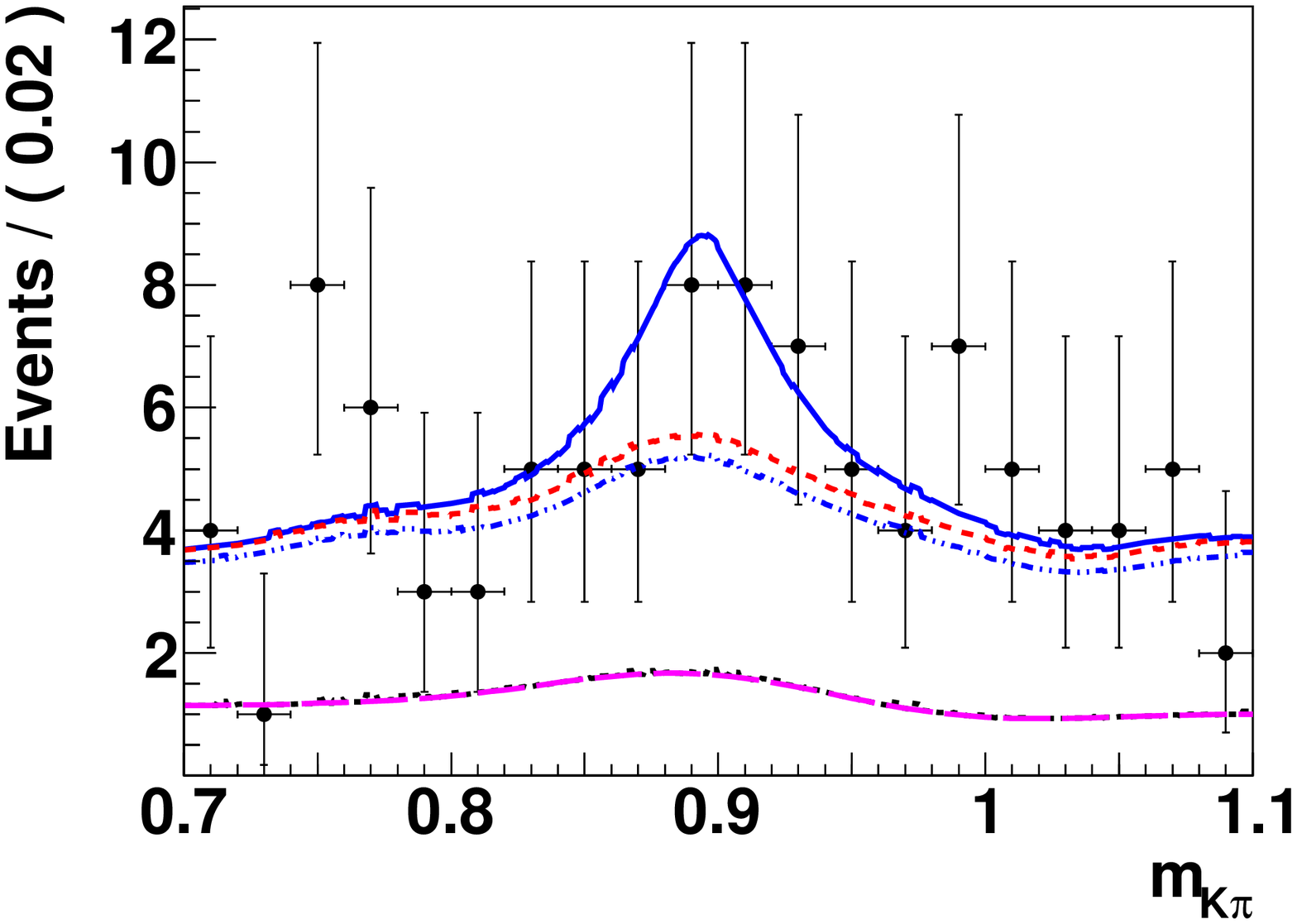}}
\subfigure[~$L_R$: \modeeight.]{\label{fig:lhrmm}\includegraphics[width=.32\textwidth]{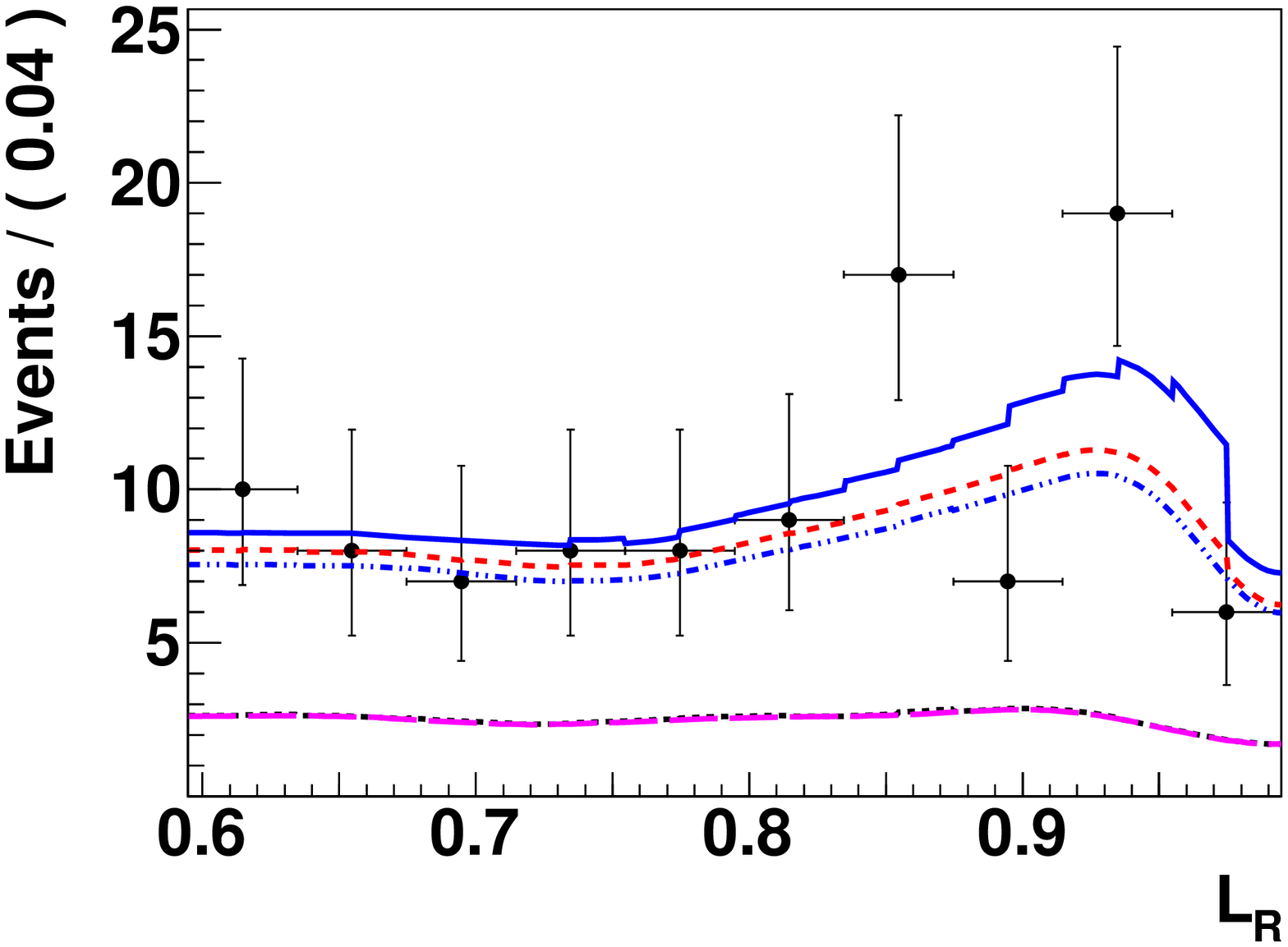}}
\caption{Initial 3-d fit projections for
$\modetwelve$ (top row) and $\modeeight$ (bottom row) in $q^{2}_5$.
  The plots show the stacked contributions from each event class:
  combinatorial (magenta long dash),
  charmonium (black dots),
  crossfeed (red short dash),
  total pdf (solid blue)
  and, in the bottom row of plots
  only, muon mis-identification (blue
  dash dots). The signal pdf is
  represented by the area between
  the dash red and solid blue lines.}
\label{fig:firstfit}
\end{figure*}

As discussed above, the initial three-dimensional fits to $\mes$, $m(K\pi)$ and $L_R$ are
done using events in the full $\mes > 5.2 \gevcc$ range;
each final state in each $q^2$ bin is separately fit
in order to establish the normalizations and pdf shapes
subsequently used in extracting the angular observables
from the $\mes > 5.27 \gevcc$ angular fit region.
Table~\ref{tab:combinedyields} gives the resulting fitted
signal yields along with statistical uncertainties for the three different
combinations of particular final states for which the angular observables are extracted.
As examples of typical fits, Fig.~\ref{fig:firstfit} shows
fit projections in each of the three initial fit dimensions for
$\modetwelve$ and $\modeeight$ in the $q^{2}_5$ bin.
Validation of the initial 3-d fit model is done using
events in the $\jpsi$ and $\psitwos$ dilepton mass veto
windows, where we find good agreement between our fit results and the
nominal PDG values for the $B \to \jpsi \Kstar$ and
$B \to \psitwos \Kstar$ branching fractions~\cite{Agashe:2014kda}
into our final states.

\begin{table*}
\centering
\caption[Fitted signal yields with statistical uncertainties from the initial 3-d fit.]
{Fitted signal yields with statistical uncertainties.}
\begin{tabular}{l @{\extracolsep{2em}} c @{\extracolsep{2em}} c @{\extracolsep{2em}} c @{\extracolsep{2em}} c @{\extracolsep{2em}} c @{\extracolsep{2em}} c}
\\ \hline \hline \\
Mode & $q^2_0$ & $q^2_1$ & $q^2_2$ & $q^2_3$ & $q^2_4$ & $q^2_5$ \\ \hline
\noalign{\vskip 1mm}
$\modekstll$   & $40.8 \pm 8.4$ & $31.7 \pm 7.1$ & $11.9 \pm 5.5$ & $21.3 \pm 8.5$ & $31.9 \pm 9.2$ & $33.2 \pm 7.8$ \\
$\modekstksll$ & $17.7 \pm 5.2$ & $8.7  \pm 4.1$ & $3.8  \pm 4.0$ & $7.7  \pm 5.6$ & $9.0  \pm 4.8$ & $9.4  \pm 4.2$ \\
$\modekstzll$  & $23.1 \pm 6.6$ & $22.9 \pm 5.8$ & $8.1  \pm 3.8$ & $13.7 \pm 6.4$ & $22.8 \pm 7.8$ & $23.8 \pm 6.6$ \\
\hline \hline
\label{tab:combinedyields}
\end{tabular}
\end{table*}

\subsection{Angular Fit Results}

Prior to fitting the $\modekstll$ angular data, we validate our
angular fit model by using it to extract the $\Kstar$ longitudinal
polarization $\fl$ for $B \to \jpsi \Kstar$ and
$B \to \psitwos \Kstar$ decays into our signal final states, and
comparing our results to previously reported PDG values~\cite{Agashe:2014kda}.
We also perform similar validation fits for $\afb$, which is expected
in the SM to approach zero for lepton pairs from $B$ decays to final states including charmonia.
Recalculating the PDG averages after removing all contributing \babar\, results,
we find no significant deviations from the
expected values in any individual final state or for the particular
combinations of final states used in our main analysis.

\begin{figure*}
\centering
\subfigure[~$\ctk$ $q^{2}_{0}$]{\label{fig:kpa}\begin{overpic}[width=.49\textwidth]{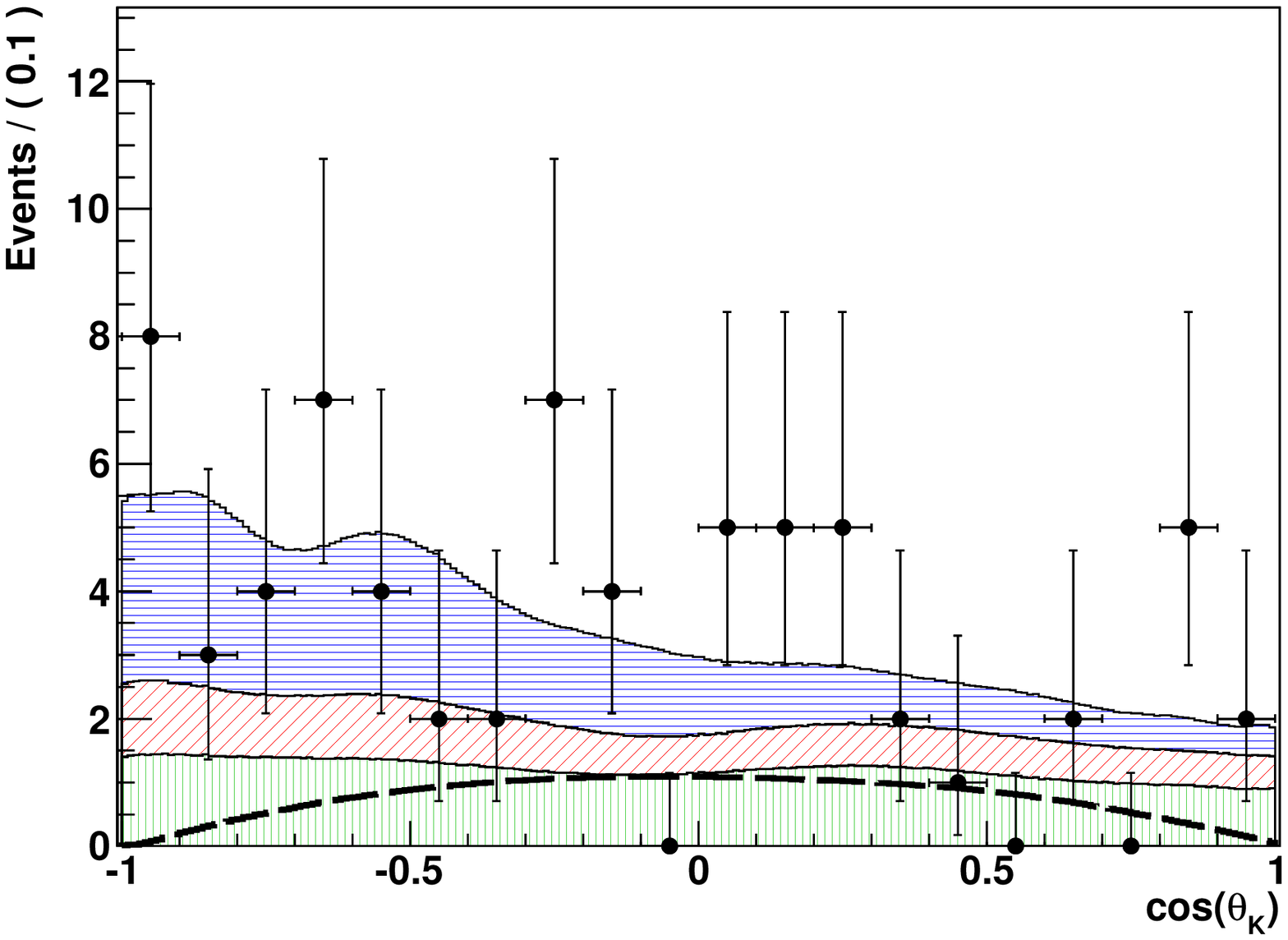}
    \put(90,2){\fboxsep=0pt\makebox[0pt][r]{\colorbox{white}{\smaller\bfseries\boldmath$\ctk$\rule[-0.6ex]{0pt}{0pt}}}}
  \end{overpic}}
\subfigure[~$\ctl$ $q^{2}_{0}$]{\label{fig:kpb}\begin{overpic}[width=.49\textwidth]{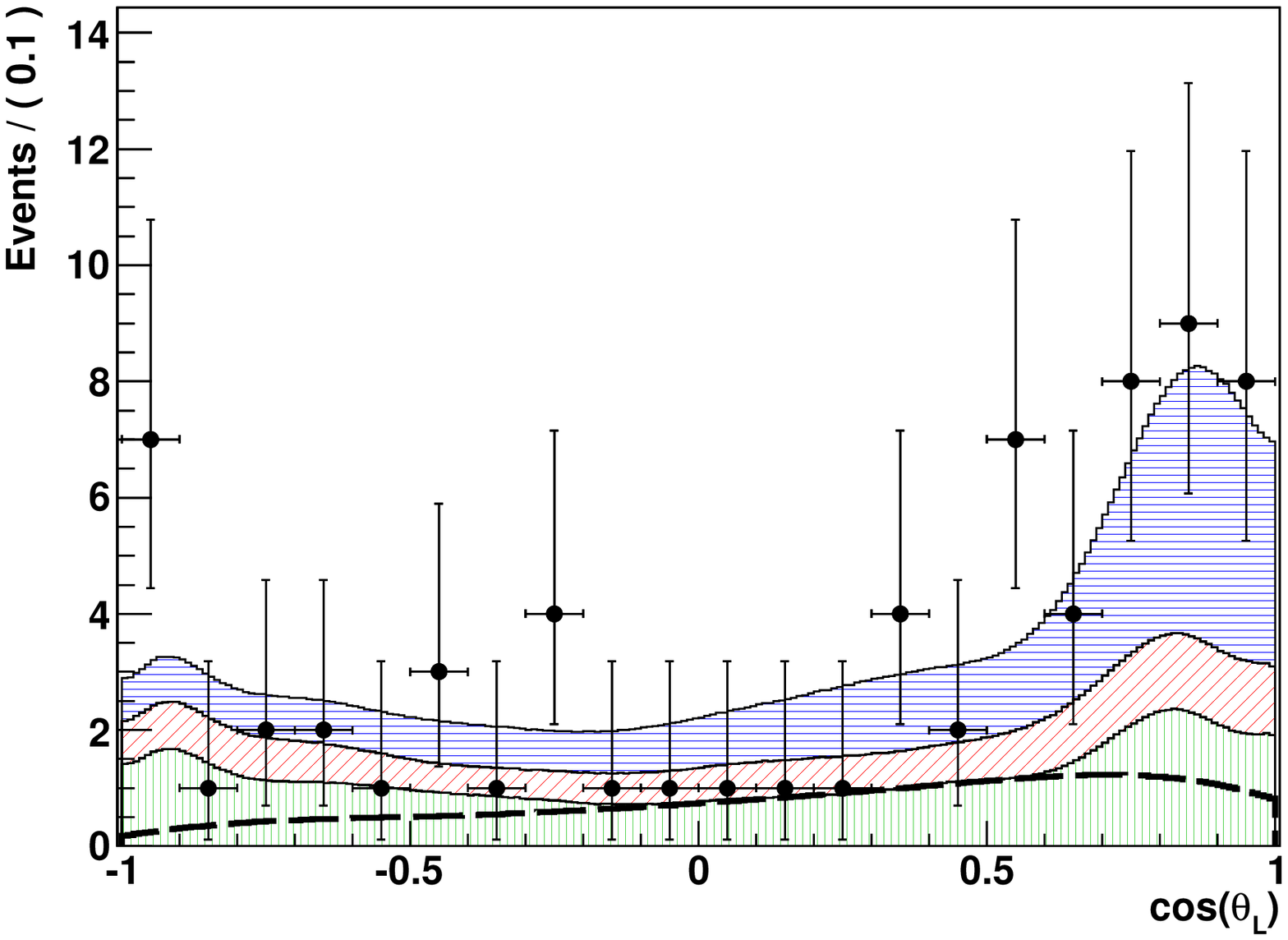}
    \put(90,2){\fboxsep=0pt\makebox[0pt][r]{\colorbox{white}{\smaller\bfseries\boldmath$\ctl$\rule[-0.6ex]{0pt}{0pt}}}}
  \end{overpic}}
\subfigure[~$\ctk$ $q^{2}_{5}$]{\label{fig:kpc}\begin{overpic}[width=.49\textwidth]{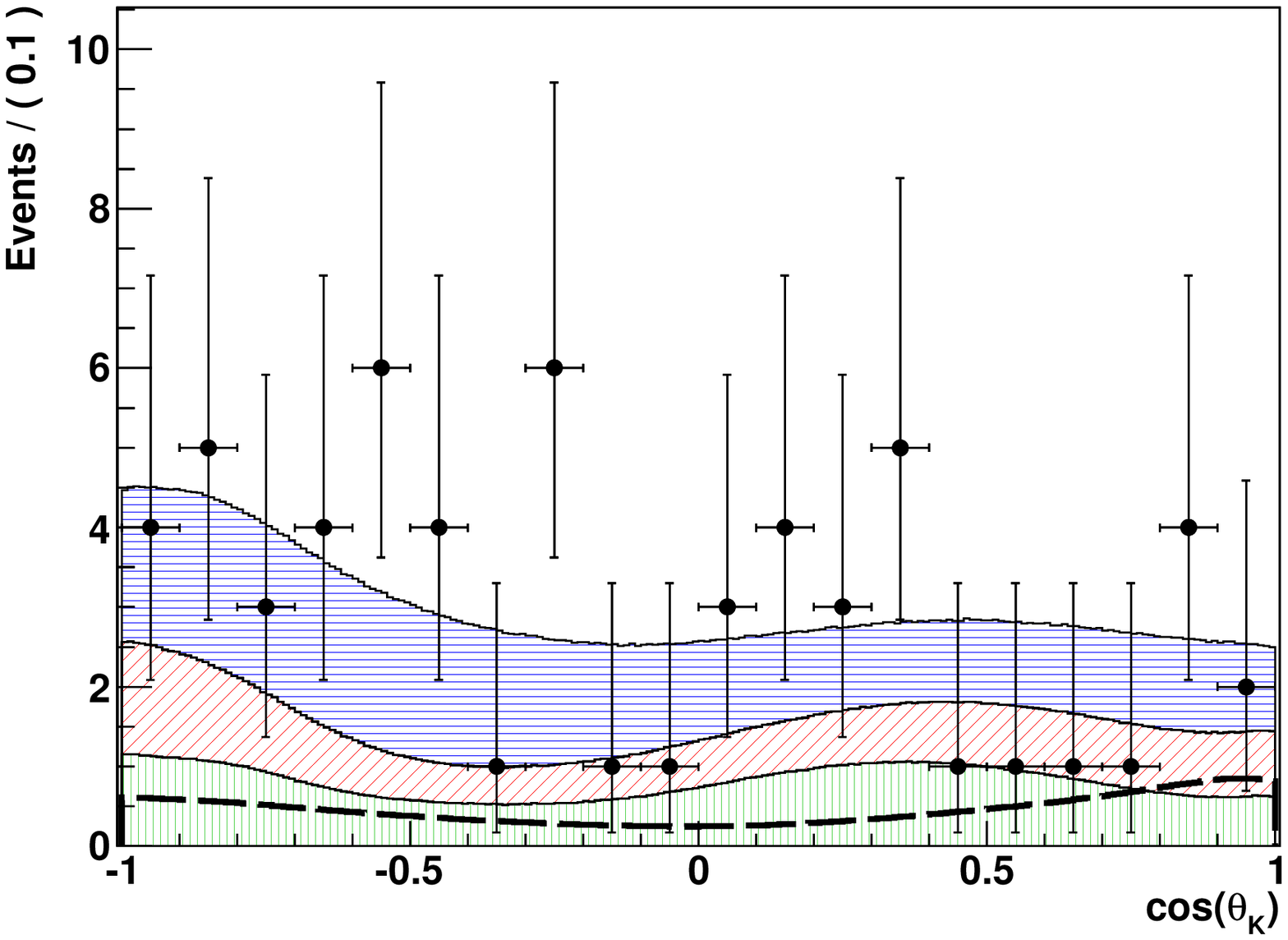}
    \put(90,2){\fboxsep=0pt\makebox[0pt][r]{\colorbox{white}{\smaller\bfseries\boldmath$\ctk$\rule[-0.6ex]{0pt}{0pt}}}}
  \end{overpic}}
\subfigure[~$\ctl$ $q^{2}_{5}$]{\label{fig:kpd}\begin{overpic}[width=.49\textwidth]{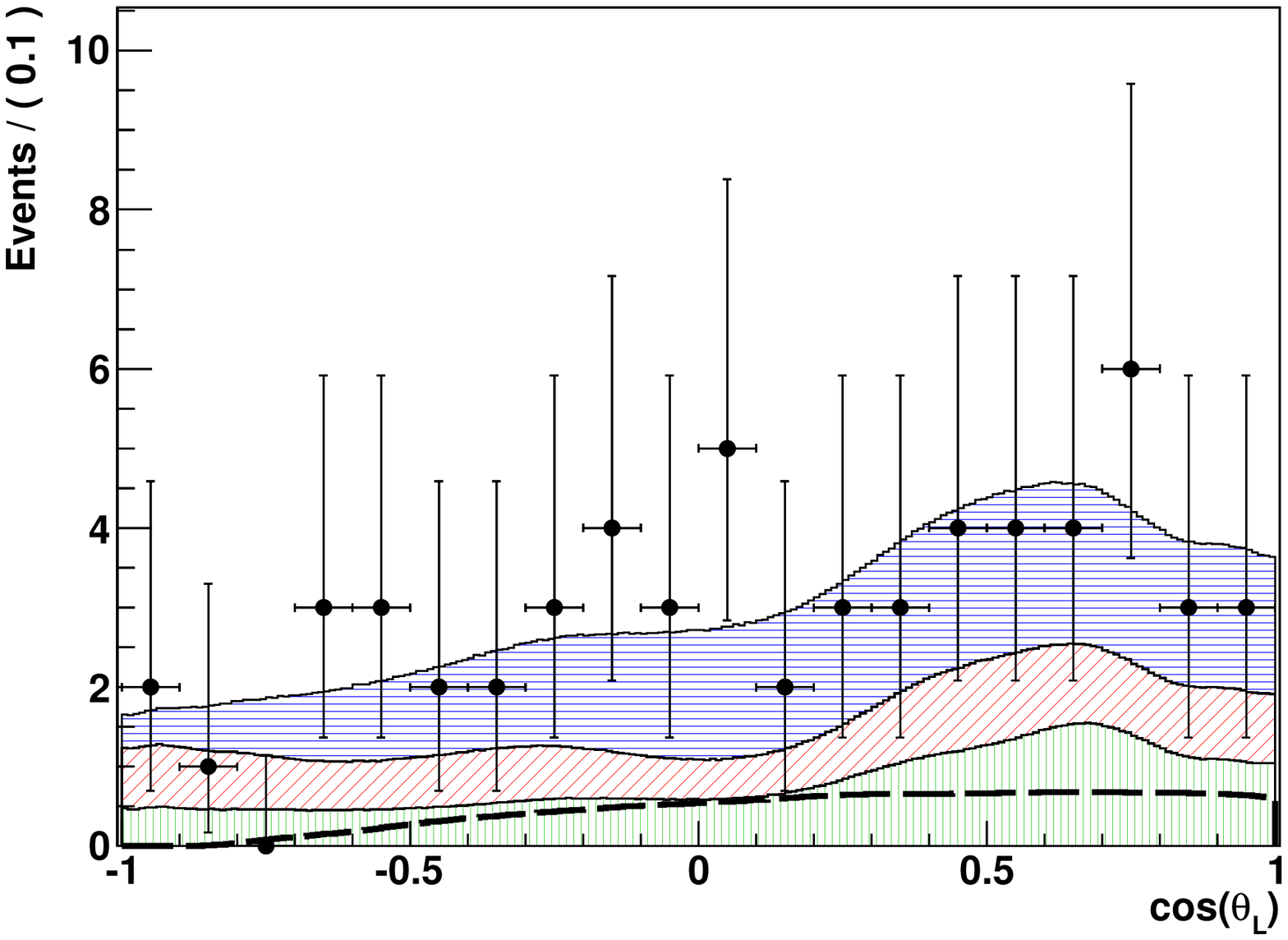}
    \put(90,2){\fboxsep=0pt\makebox[0pt][r]{\colorbox{white}{\smaller\bfseries\boldmath$\ctl$\rule[-0.6ex]{0pt}{0pt}}}}
  \end{overpic}}
\caption{$\modekstksll$ angular fit projections.
The shaded areas show the contribution to the
total fit from each individual final state:
(green vertical lines) $\modeseven$;
(red diagonal lines) $\modeeleven$;
(blue horizontal lines) $\modeten$.
The overlaid dashed line shows the total signal contribution summed over the three individual final states.
Each colored band includes both signal and background events in a given final state.}
\label{fig:flafbfitsplus}
\end{figure*}

\begin{figure*}
\centering
\subfigure[~$\ctk$ $q^{2}_{0}$]{\label{fig:kza}\begin{overpic}[width=.49\textwidth]{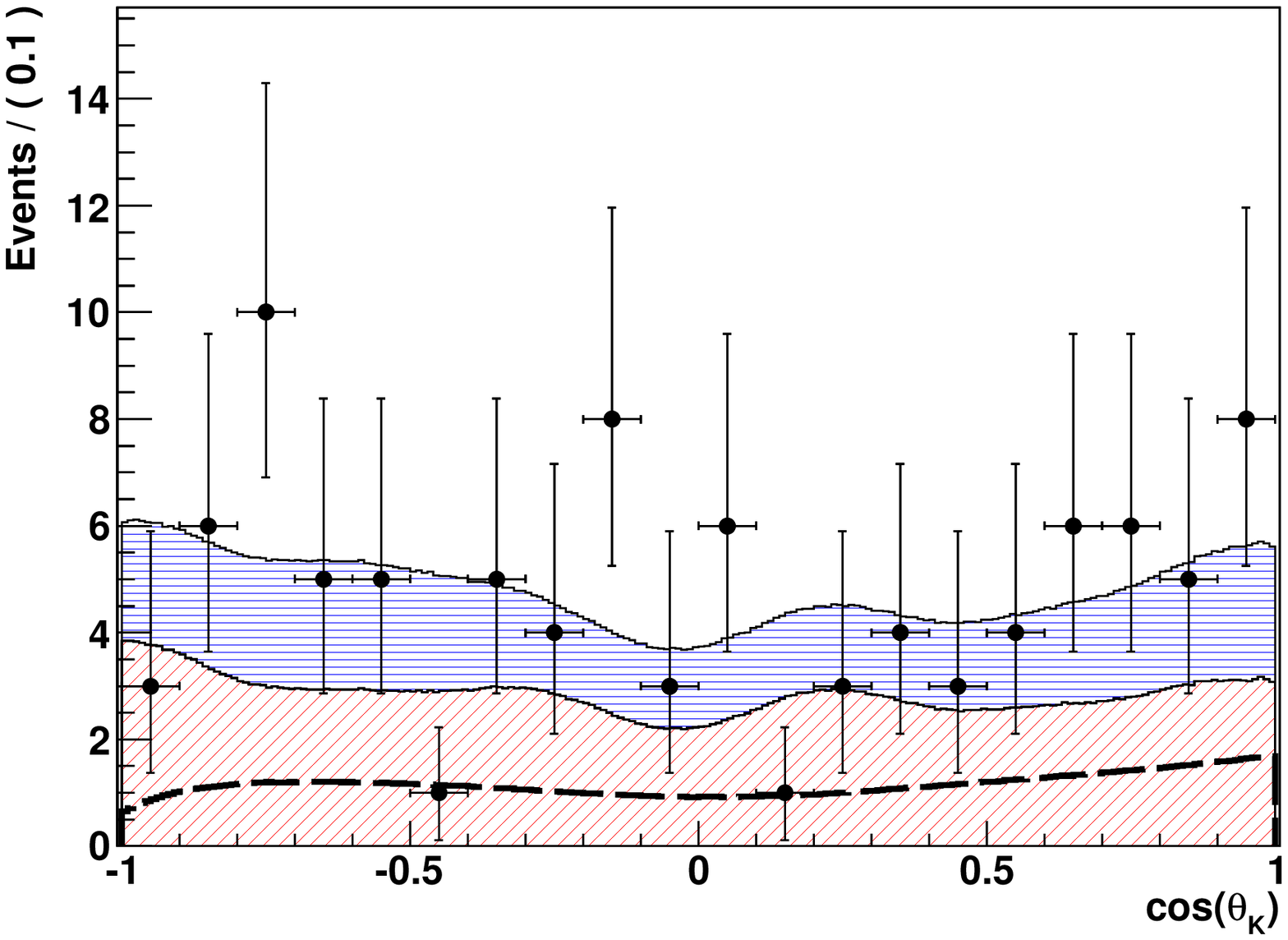}
    \put(90,2){\fboxsep=0pt\makebox[0pt][r]{\colorbox{white}{\smaller\bfseries\boldmath$\ctk$\rule[-0.6ex]{0pt}{0pt}}}}
  \end{overpic}}
\subfigure[~$\ctl$ $q^{2}_{0}$]{\label{fig:kzb}\begin{overpic}[width=.49\textwidth]{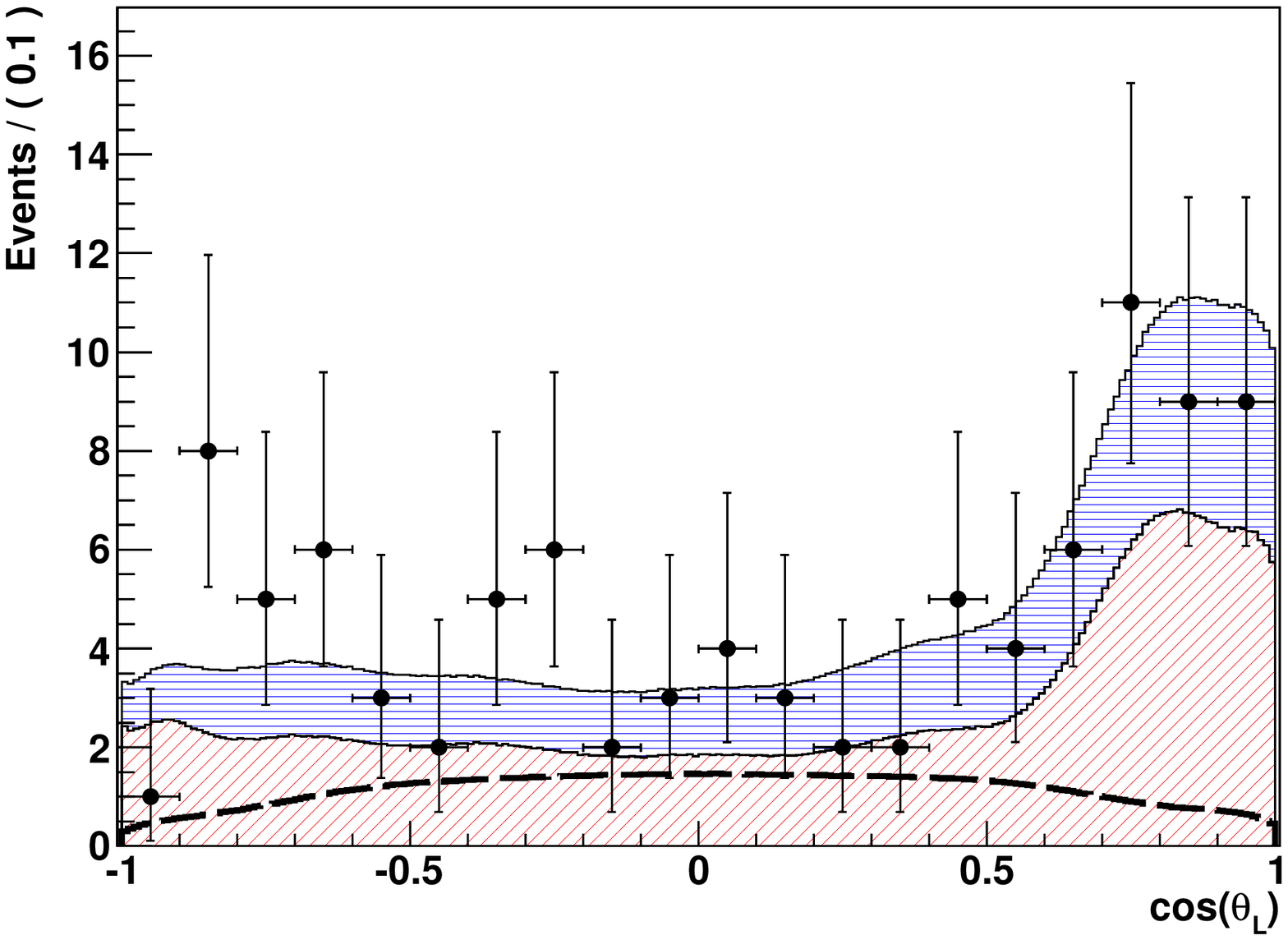}
    \put(90,2){\fboxsep=0pt\makebox[0pt][r]{\colorbox{white}{\smaller\bfseries\boldmath$\ctl$\rule[-0.6ex]{0pt}{0pt}}}}
  \end{overpic}}
\subfigure[~$\ctk$ $q^{2}_{5}$]{\label{fig:kzc}\begin{overpic}[width=.49\textwidth]{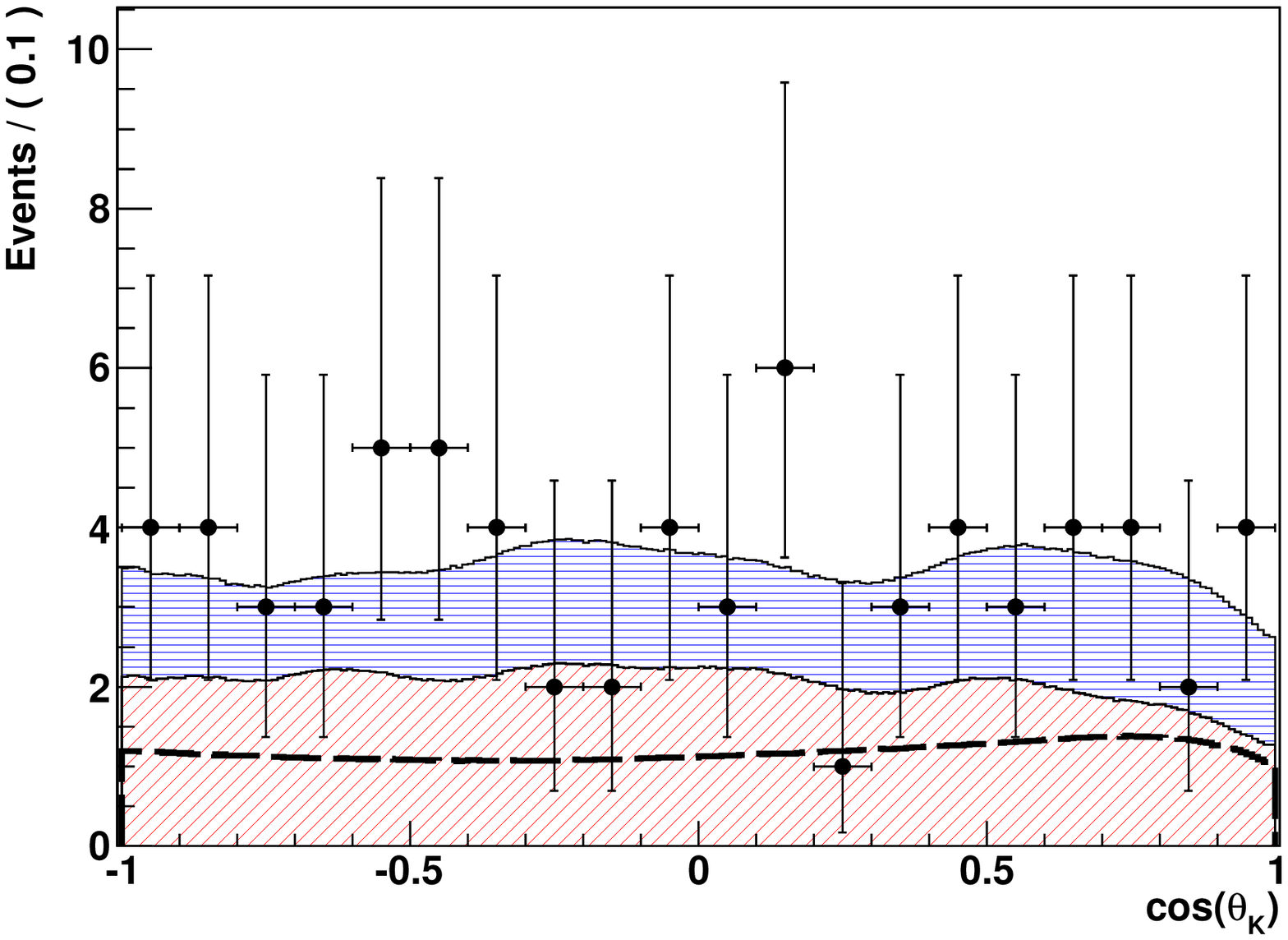}
    \put(90,2){\fboxsep=0pt\makebox[0pt][r]{\colorbox{white}{\smaller\bfseries\boldmath$\ctk$\rule[-0.6ex]{0pt}{0pt}}}}
  \end{overpic}}
\subfigure[~$\ctl$ $q^{2}_{5}$]{\label{fig:kzd}\begin{overpic}[width=.49\textwidth]{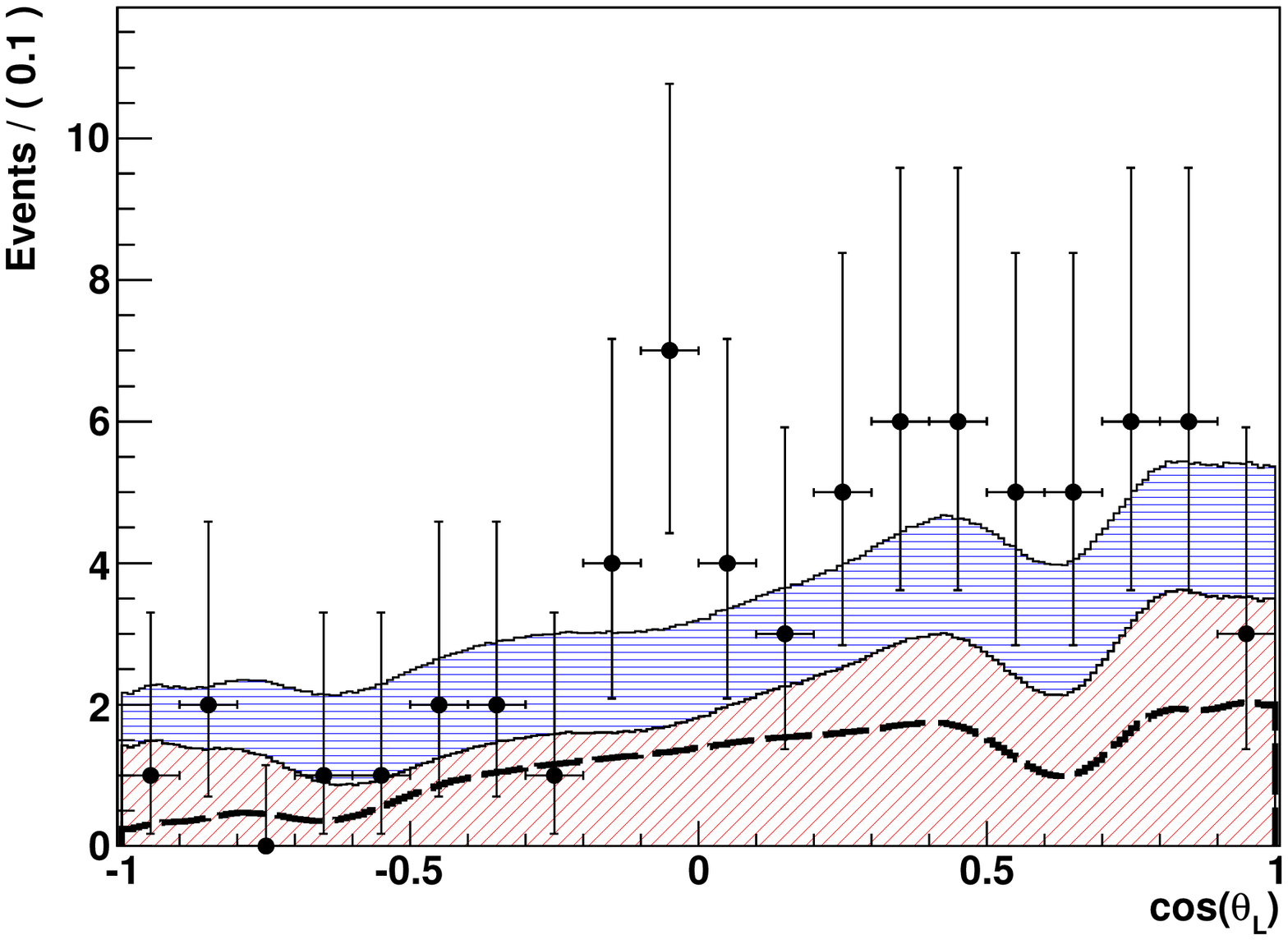}
    \put(90,2){\fboxsep=0pt\makebox[0pt][r]{\colorbox{white}{\smaller\bfseries\boldmath$\ctl$\rule[-0.6ex]{0pt}{0pt}}}}
  \end{overpic}}
\caption{$\modekstzll$ angular fit projections.
The shaded areas show the contribution to the
total fit from each individual final state:
(red diagonal lines) $\modeeight$;
(blue horizontal lines) $\modetwelve$.
The overlaid dashed line shows the total signal contribution summed over the two individual final states.
Each colored band includes both signal and background events in a given final state.}
\label{fig:flafbfitszero}
\end{figure*}

\begin{figure*}
\centering
\subfigure[~$\fl$.]{\label{fig7a}\begin{overpic}[width=.95\textwidth]{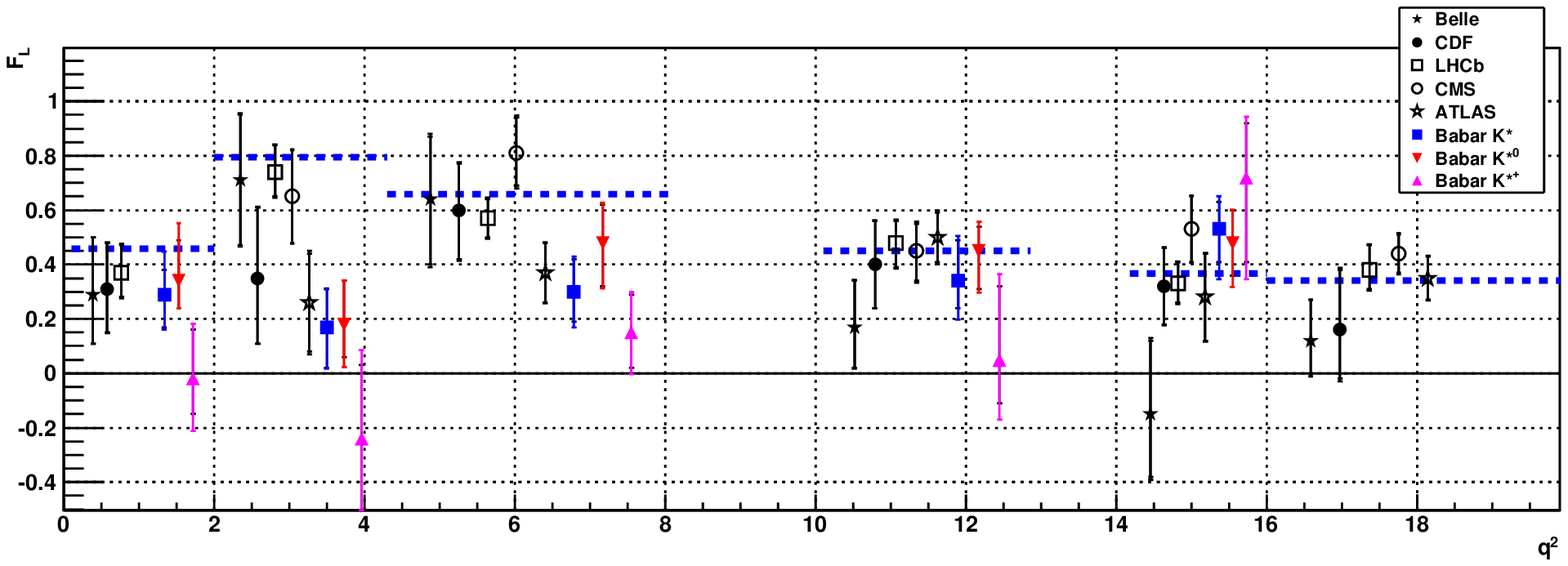}
    \put(2.5,34){%
      \begin{rotate}{90}
        \makebox[0pt][r]{\colorbox{white}{\smaller\bfseries\boldmath$\fl$}\rule[-0.6ex]{0.5em}{0pt}}
      \end{rotate}}
  \end{overpic}}
\subfigure[~$\afb$.]{\label{fig7b}\begin{overpic}[width=.95\textwidth]{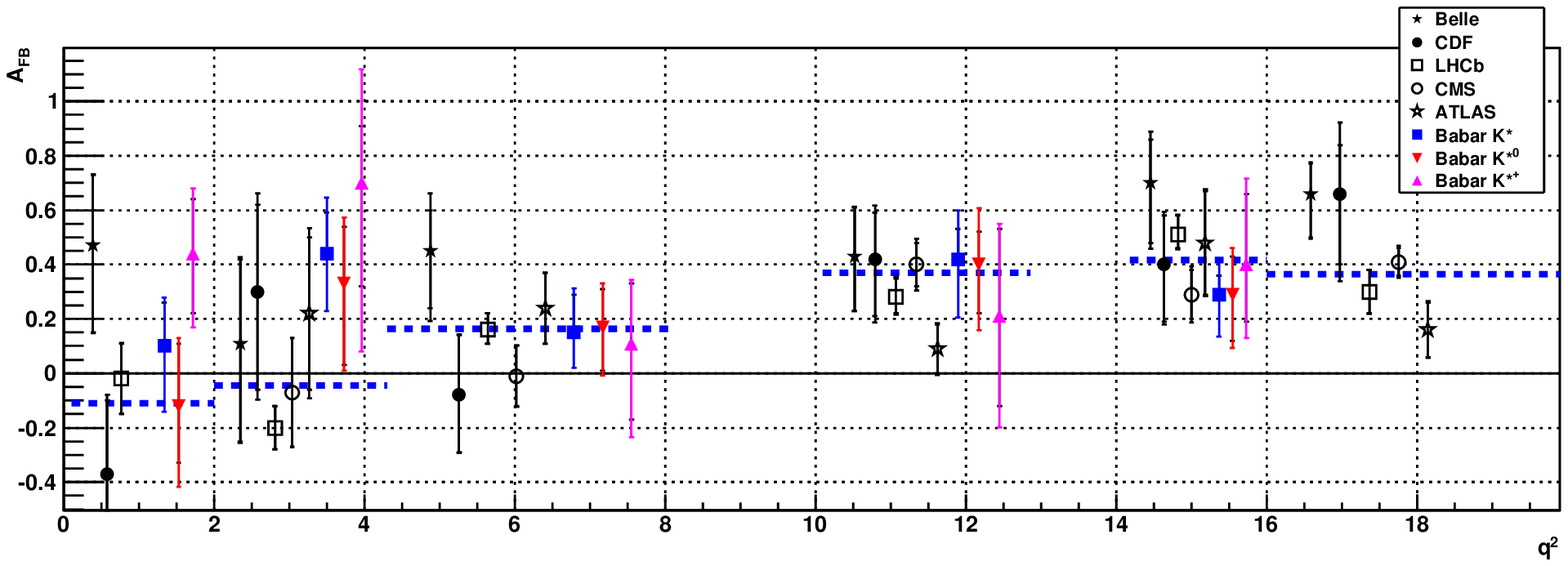}
    \put(2.5,34){%
      \begin{rotate}{90}
        \makebox[0pt][r]{\colorbox{white}{\smaller\bfseries\boldmath$\afb$}\rule[-0.6ex]{0.5em}{0pt}}
      \end{rotate}}
  \end{overpic}}
\caption{$\fl$ (top) and $\afb$ (bottom) results in disjoint $q^2$ bins, along with those of other experiments and the SM expectations (blue dashed lines,
which also define the extent of each individual $q^2$ bin):
(black filled star) Belle~\cite{Wei:2009zv},
(black filled circle) CDF~\cite{Aaltonen:2011ja},
(black open square) LHCb~\cite{Aaij:2013qta},
(black open circle) CMS~\cite{Chatrchyan:2013cda},
(black open star) ATLAS~\cite{ATLAS:2013ola},
(blue filled square) \babar\, $\modekstll$,
(red filled down-pointing triangle) $\modekstzll$ ,
(magenta filled up-pointing triangle) $\modekstksll$.
The \babar\, $q^2_5$ results are drawn in the
$14 \lesssim q^2 < 16 \gevcccc$ region, however, they are
valid for the entire $q^2 \gtrsim 14 \gevcccc$ region.}
\label{fig:allresults}
\end{figure*}

\begin{figure*}
\centering
\subfigure[~$\fl$.]{\label{fig8a}\includegraphics[width=.45\textwidth]{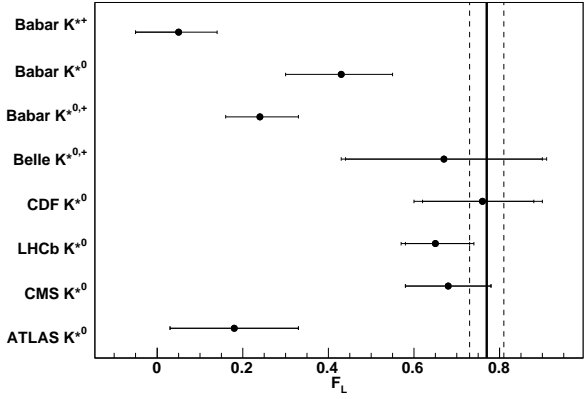}}
\subfigure[~$\afb$.]{\label{fig8b}\includegraphics[width=.45\textwidth]{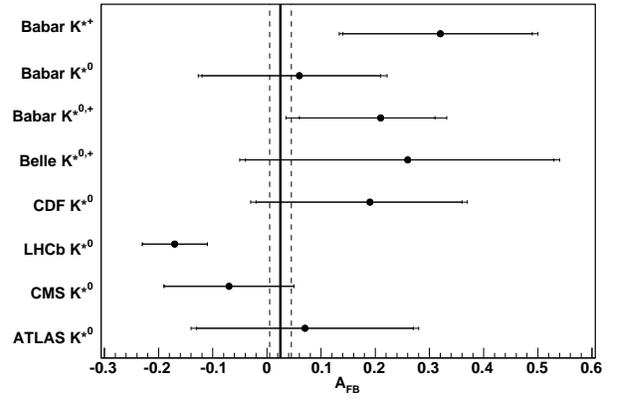}}
\caption{$q^2_0$ $\fl$ (left) and $\afb$ (right) results, along with those of other
experiments~\cite{Wei:2009zv,Aaltonen:2011ja,Aaij:2013qta,Chatrchyan:2013cda,ATLAS:2013ola}
and the SM expectation (vertical lines)~\cite{Buchalla:1995vs,AFB_SMa,AFB_SMb,AFB_SMc,Altmannshofer:2008dz,Hovhannisyan}.}
\label{fig:q0results}
\end{figure*}

Having validated our fit model with the vetoed charmonium events, we proceed
to the extraction of the angular observables in each $q^2$ bin.
Our results are tabulated in Tables~\ref{tab:flresults} and~\ref{tab:afbresults};
Figs.~\ref{fig:flafbfitsplus} and~\ref{fig:flafbfitszero} show the
$\modekstksll$ and $\modekstzll$ $\ctk$ and $\ctl$ fit projections
in $q^{2}_{0}$ and $q^{2}_{5}$.
Fig.~\ref{fig:allresults} graphically shows our $\fl$ and $\afb$ results in disjoint $q^2$ bins
alongside other published results and the SM theory expectations, the latter of which
typically have 5-10\% theory uncertainties (absolute) in the regions below and above
the charmonium resonances.
Fig.~\ref{fig:q0results} similarly compares the $q^2_0$ results obtained here with those of other
experiments and the SM theory expectation.

\begin{table*}
\centering
\caption[$\fl$ angular fit results with, respectively, statistical and systematic uncertainties.]
{$\fl$ angular fit results with, respectively, statistical and systematic uncertainties.}
\begin{tabular}{l @{\extracolsep{4em}} c @{\extracolsep{4em}} c @{\extracolsep{4em}} c}
\\ \hline \hline \\
                      & $\modekstksll$                                & $\modekstzll$                                 & $\modekstll$ \\ \hline \noalign{\vskip 1mm}
$q^2_0$               & $+0.05_{-0.10}^{+0.09}{}_{-0.10}^{+0.02}$ & $+0.43_{-0.13}^{+0.12}{}_{-0.02}^{+0.02}$ & $+0.24_{-0.08}^{+0.09}{}_{-0.02}^{+0.02}$ \\ \noalign{\vskip 1mm}
$q^2_1$               & $-0.02_{-0.13}^{+0.18}{}_{-0.14}^{+0.09}$ & $+0.34_{-0.10}^{+0.15}{}_{-0.02}^{+0.15}$ & $+0.29_{-0.12}^{+0.09}{}_{-0.05}^{+0.13}$ \\ \noalign{\vskip 1mm}
$q^2_2$               & $-0.24_{-0.39}^{+0.27}{}_{-0.10}^{+0.18}$ & $+0.18_{-0.12}^{+0.16}{}_{-0.10}^{+0.02}$ & $+0.17_{-0.15}^{+0.14}{}_{-0.02}^{+0.02}$ \\ \noalign{\vskip 1mm}
$q^2_3$               & $+0.15_{-0.13}^{+0.14}{}_{-0.08}^{+0.05}$ & $+0.48_{-0.16}^{+0.14}{}_{-0.05}^{+0.05}$ & $+0.30_{-0.11}^{+0.12}{}_{-0.07}^{+0.05}$ \\ \noalign{\vskip 1mm}
$q^2_4$               & $+0.05_{-0.16}^{+0.27}{}_{-0.15}^{+0.16}$ & $+0.45_{-0.14}^{+0.09}{}_{-0.06}^{+0.06}$ & $+0.34_{-0.10}^{+0.15}{}_{-0.10}^{+0.07}$ \\ \noalign{\vskip 1mm}
$q^2_5$               & $+0.72_{-0.31}^{+0.20}{}_{-0.21}^{+0.10}$ & $+0.48_{-0.12}^{+0.12}{}_{-0.11}^{+0.02}$ & $+0.53_{-0.12}^{+0.10}{}_{-0.14}^{+0.07}$ \\ \noalign{\vskip 1mm}
\hline \hline
\label{tab:flresults}
\end{tabular}
\end{table*}

\begin{table*}
\centering
\caption[$\afb$ angular fit results with, respectively, statistical and systematic uncertainties.]
{$\afb$ angular fit results with, respectively, statistical and systematic uncertainties.}
\begin{tabular}{l @{\extracolsep{4em}} c @{\extracolsep{4em}} c @{\extracolsep{4em}} c}
\\ \hline \hline \\
                      & $\modekstksll$                                 & $\modekstzll$                                  & $\modekstll$ \\ \hline \noalign{\vskip 1mm}
$q^2_0$               & $+0.32_{-0.18}^{+0.18}{}_{-0.05}^{+0.08}$ & $+0.06_{-0.18}^{+0.15}{}_{-0.05}^{+0.06}$ & $+0.21_{-0.15}^{+0.10}{}_{-0.09}^{+0.07}$ \\ \noalign{\vskip 1mm}
$q^2_1$               & $+0.44_{-0.22}^{+0.20}{}_{-0.16}^{+0.13}$ & $-0.12_{-0.21}^{+0.23}{}_{-0.21}^{+0.10}$ & $+0.10_{-0.15}^{+0.16}{}_{-0.19}^{+0.08}$ \\ \noalign{\vskip 1mm}
$q^2_2$               & $+0.70_{-0.38}^{+0.21}{}_{-0.49}^{+0.36}$ & $+0.33_{-0.30}^{+0.21}{}_{-0.11}^{+0.12}$ & $+0.44_{-0.18}^{+0.15}{}_{-0.11}^{+0.14}$ \\ \noalign{\vskip 1mm}
$q^2_3$               & $+0.11_{-0.28}^{+0.22}{}_{-0.20}^{+0.08}$ & $+0.17_{-0.16}^{+0.14}{}_{-0.08}^{+0.08}$ & $+0.15_{-0.12}^{+0.14}{}_{-0.05}^{+0.08}$ \\ \noalign{\vskip 1mm}
$q^2_4$               & $+0.21_{-0.33}^{+0.32}{}_{-0.24}^{+0.11}$ & $+0.40_{-0.18}^{+0.12}{}_{-0.16}^{+0.17}$ & $+0.42_{-0.17}^{+0.11}{}_{-0.13}^{+0.14}$ \\ \noalign{\vskip 1mm}
$q^2_5$               & $+0.40_{-0.21}^{+0.26}{}_{-0.17}^{+0.18}$ & $+0.29_{-0.17}^{+0.14}{}_{-0.10}^{+0.10}$ & $+0.29_{-0.10}^{+0.07}{}_{-0.12}^{+0.10}$ \\ \noalign{\vskip 1mm}
\hline \hline
\label{tab:afbresults}
\end{tabular}
\end{table*}

\subsection{Systematic Uncertainties}

We describe below the systematic uncertainties in the angular results arising from:
\begin{itemize}
\item the purely statistical uncertainties in the parameters obtained from the
initial 3-d $\mes,m(K\pi)$ fit which are used in the angular fits;
\item the $\fl$ statistical uncertainty, which is propagated into the $\afb$ fit; and
\item the modeling of the random combinatorial background pdfs and the signal angular efficiencies.
\end{itemize}

We additionally examined several other possible sources of systematic
uncertainty, but found no significant contributions due to:
\begin{itemize}
\item modeling of the signal crossfeed contributions to the angular fits;
\item the parameterization of the signal Gaussian $\mes$ and resonant $m(K\pi)$ shapes
that are extracted from the relatively high-statistics $\jpsi$ control samples;
\item possible fit biases which, to relatively very good precision, were not
observed in any of the data control sample angular fits;
\item characterization of $\mes$ peaking backgrounds from
muon mis-identification and charmonium leakage;
\item variations in event selection.
\end{itemize}

We combine in quadrature the individual systematic uncertainties
to obtain the total systematic uncertainty on each of the
angular observables; these are are given in Table~\ref{tab:totalsys},
which is placed after the detailed discussion below for each family of
systematic uncertainties.

In the initial fits that determine the signal yields, we allow the
random combinatorial $\mes$ shape and normalisation, as well as the
signal yield, to float. We then fix these parameters at their
central values for the angular fits. To study the systematic
uncertainty associated with these fixed parameters, we vary each
parameter from its central value by its $\pm 1\sigma$ statistical
uncertainty, accounting for correlations among the fit parameters,
and then redo the angular fit. To control for systematic fit results
that deviate from the nominal central value mainly from statistical
effects rather than systematic ones, we additionally examine fit results obtained
from $\pm (0.8,0.9,1.1,1.2) \sigma$ variations. These small
variations on the $\pm 1\sigma$ values should also result in
similarly small variations, in the absence of any statistical effects, on a
$\pm 1\sigma$ systematic fit result. For the bulk of the systematics,
where the series of fit results for each of the additional variations
is linearly distributed around the middle $1\sigma$ fit result, the
$1\sigma$ variation is considered robust.
In the relatively few cases where the disagreement between the
nominal $1\sigma$ variation and the value
of the $1\sigma$ variation interpolated from the additional
$(0.8,0.9,1.1,1.2) \sigma$ variations is statistically significant,
the interpolated $1\sigma$ value is used to assign the systematic.
All deviations from the nominal fit central value are then added in
quadrature to obtain the overall systematic uncertainty attributable to this
source, which is given in Table~\ref{tab:floatparmsys}.

The $\ctk$ fit yields the central value and statistical uncertainty
for $\fl$ in each $q^2$ bin, which is subsequently used in the fit
to the $\ctl$ distributions to extract $\afb$. To study the systematic
uncertainty on $\afb$ due to the purely statistical $\fl$ uncertainty,
we vary the value of $\fl$ by $\pm 1\sigma$ from its fitted value, and
redo the $\ctl$ fits with the new value of $\fl$. We determine the
systematic uncertainty from the shift in the central value of $\afb$
relative to the nominal fit for $\pm 1\sigma$ variations of $\fl$;
these are given in Table~\ref{tab:flinafbsys}. As with the variations
described in the preceding paragraph, additional fits for several $\fl$
variations surrounding the nominal $\pm 1\sigma$ values are performed.
We then apply the same quality criterion as for the preceding systematic and,
where this criterion is not met, assign the $\fl$ systematic using an
interpolated $1\sigma$ value rather than the fitted $1\sigma$ variation.

The angular combinatorial background shapes are derived from the $\mes$ sideband
region and are non-parametrically modeled directly from these data. We examine
several variations on the modeling, and additionally use the LFV events (described
above) as an alternative dataset from which the angular background pdfs are drawn.
We assign a systematic uncertainty associated with the modeling of
these pdfs by using 20 different variations of the non-parametric
modeling and refitting for $\fl$ and $\afb$.
We take the largest of the deviations between the default nominal fit and
these varied fit results, and to this add in quadrature the deviation from the nominal fit obtained using
the LFV dataset; the resulting systematic uncertainty is given in Table~\ref{tab:lfvsys}.

Finally, to study a possible systematic uncertainty on $\fl$ and $\afb$ as a function
of their true physical values, we generated and reconstructed simulated
events with varied values of the underlying Wilson coefficients $C_7$, $C_9$, and $C_{10}$
in order to produce a range of near-maximal, but physically allowed, asymmetries.
These datasets are used to produce signal efficiency histograms differing
from the default ones, which use the expected SM values for the Wilson coefficients. This allows
different regions of the angular distributions to contribute with
different relative weight depending on the magnitude and sign of the underlying angular asymmetries. Applying these
alternative signal efficiency histograms, we measure the shifts in the
fitted values of $\fl$ and $\afb$, and assign as the systematic the sum-in-quadrature of each deviation
from the nominal central value; the resulting systematic uncertainty is shown in Table~\ref{tab:genvarsys}.

\begin{table*}
\centering
\caption{Angular observable systematic uncertainties from the initial 3-d fit.}
\begin{tabular}{l @{\extracolsep{1em}} c @{\extracolsep{1em}} c @{\extracolsep{1em}} c @{\extracolsep{1em}} c @{\extracolsep{1em}} c @{\extracolsep{1em}} c}
\\ \hline \hline \\
        & \multicolumn{3}{c}{$\fl$ systematic}         & \multicolumn{3}{c}{$\afb$ systematic} \\
\noalign{\vskip 1mm}         & $\modekstksll$  & $\modekstzll$   & $\modekstll$    & $\modekstksll$  & $\modekstzll$   & $\modekstll$      \\ \hline
\noalign{\vskip 1mm} $q^2_0$ & $+0.02$ $-0.09$ & $+0.02$ $-0.02$ & $+0.02$ $-0.02$ & $+0.05$ $-0.04$ & $+0.01$ $-0.04$ & $+0.02$ $-0.07$     \\
\noalign{\vskip 1mm} $q^2_1$ & $+0.09$ $-0.13$ & $+0.02$ $-0.02$ & $+0.02$ $-0.05$ & $+0.12$ $-0.08$ & $+0.07$ $-0.02$ & $+0.07$ $-0.08$     \\
\noalign{\vskip 1mm} $q^2_2$ & $+0.18$ $-0.05$ & $+0.02$ $-0.01$ & $+0.02$ $-0.02$ & $+0.34$ $-0.48$ & $-0.02$ $-0.08$ & $+0.09$ $-0.07$     \\
\noalign{\vskip 1mm} $q^2_3$ & $+0.05$ $-0.07$ & $+0.02$ $-0.02$ & $+0.05$ $-0.06$ & $+0.02$ $-0.19$ & $-0.02$ $-0.04$ & $+0.01$ $-0.02$     \\
\noalign{\vskip 1mm} $q^2_4$ & $+0.11$ $-0.14$ & $+0.02$ $-0.06$ & $+0.02$ $-0.10$ & $+0.09$ $-0.23$ & $+0.15$ $-0.11$ & $+0.13$ $-0.10$     \\
\noalign{\vskip 1mm} $q^2_5$ & $+0.02$ $-0.19$ & $+0.02$ $-0.10$ & $+0.02$ $-0.14$ & $+0.16$ $-0.09$ & $+0.05$ $-0.02$ & $+0.08$ $-0.02$      \\
\noalign{\vskip 1mm}
\hline \hline
\end{tabular}
\label{tab:floatparmsys}
\end{table*}

\begin{table*}
\centering
\caption{Systematic uncertainty in $\afb$ from the experimental determination of $\fl$.}
\begin{tabular}{l @{\extracolsep{1em}} c @{\extracolsep{1em}} c @{\extracolsep{1em}} c}
\\ \hline \hline \\
        & \multicolumn{3}{c}{$\afb$ systematic} \\
\noalign{\vskip 1mm}         & $\modekstksll$ & $\modekstzll$ & $\modekstll$ \\ \hline
\noalign{\vskip 1mm} $q^2_0$ & $\pm 0.04$     & $\pm 0.04$    & $\pm 0.04$   \\
\noalign{\vskip 1mm} $q^2_1$ & $\pm 0.04$     & $\pm 0.07$    & $\pm 0.04$   \\
\noalign{\vskip 1mm} $q^2_2$ & $\pm 0.07$     & $\pm 0.07$    & $\pm 0.08$   \\
\noalign{\vskip 1mm} $q^2_3$ & $\pm 0.03$     & $\pm 0.06$    & $\pm 0.04$   \\
\noalign{\vskip 1mm} $q^2_4$ & $\pm 0.04$     & $\pm 0.07$    & $\pm 0.06$   \\
\noalign{\vskip 1mm} $q^2_5$ & $\pm 0.08$     & $\pm 0.07$    & $\pm 0.07$   \\
\noalign{\vskip 1mm}
\hline \hline
\end{tabular}
\label{tab:flinafbsys}
\end{table*}

\begin{table*}
\centering
\caption{Systematic uncertainties from combinatorial background modeling. ``---'' denotes
where there is no uncertainty associated with a particular systematic.}
\begin{tabular}{l @{\extracolsep{1em}} c @{\extracolsep{1em}} c @{\extracolsep{1em}} c @{\extracolsep{1em}} c @{\extracolsep{1em}} c @{\extracolsep{1em}} c}
\\ \hline \hline \\
        & \multicolumn{3}{c}{$\fl$ systematic}         & \multicolumn{3}{c}{$\afb$ systematic} \\
\noalign{\vskip 1mm}         & $\modekstksll$ & $\modekstzll$ & $\modekstll$ & $\modekstksll$ & $\modekstzll$ & $\modekstll$    \\ \hline
\noalign{\vskip 1mm} $q^2_0$ & \phantom{-0.}---\phantom{0}   $-0.05$ & \phantom{-0.}---\phantom{0} \phantom{-0.}---\phantom{0}    & \phantom{-0.}---\phantom{0}  \phantom{-0.}---\phantom{0}  & $+0.04$   \phantom{-0.}---\phantom{0}   & \phantom{-0.}---\phantom{0}    \phantom{-0.}---\phantom{0}   &  \phantom{-0.}---\phantom{0} $-0.04$  \\
\noalign{\vskip 1mm} $q^2_1$ &$+0.02$ $-0.02$ & \phantom{-0.}---\phantom{0} \phantom{-0.}---\phantom{0}    & \phantom{-0.}---\phantom{0}  \phantom{-0.}---\phantom{0}  & $+0.05$   \phantom{-0.}---\phantom{0}   & \phantom{-0.}---\phantom{0}    \phantom{-0.}---\phantom{0}   &  \phantom{-0.}---\phantom{0}  \phantom{-0.}---\phantom{0}    \\
\noalign{\vskip 1mm} $q^2_2$ & \phantom{-0.}---\phantom{0}   $-0.05$ & \phantom{-0.}---\phantom{0} \phantom{-0.}---\phantom{0}    & \phantom{-0.}---\phantom{0}  \phantom{-0.}---\phantom{0}  &  \phantom{-0.}---\phantom{0}   $-0.07$  & \phantom{-0.}---\phantom{0}   $-0.04$ &  \phantom{-0.}---\phantom{0}  \phantom{-0.}---\phantom{0}    \\
\noalign{\vskip 1mm} $q^2_3$ & \phantom{-0.}---\phantom{0}     \phantom{-0.}---\phantom{0}  & \phantom{-0.}---\phantom{0} \phantom{-0.}---\phantom{0}    & \phantom{-0.}---\phantom{0}  \phantom{-0.}---\phantom{0}  &  \phantom{-0.}---\phantom{0}     \phantom{-0.}---\phantom{0}   & \phantom{-0.}---\phantom{0}    \phantom{-0.}---\phantom{0}   &  \phantom{-0.}---\phantom{0}  \phantom{-0.}---\phantom{0}    \\
\noalign{\vskip 1mm} $q^2_4$ &$+0.10$   \phantom{-0.}---\phantom{0}  & \phantom{-0.}---\phantom{0} \phantom{-0.}---\phantom{0}    & \phantom{-0.}---\phantom{0}  \phantom{-0.}---\phantom{0}  &   \phantom{-0.}---\phantom{0}  $-0.04$  & \phantom{-0.}---\phantom{0}    \phantom{-0.}---\phantom{0}   &  \phantom{-0.}---\phantom{0}  \phantom{-0.}---\phantom{0}    \\
\noalign{\vskip 1mm} $q^2_5$ & \phantom{-0.}---\phantom{0}   $-0.10$ & \phantom{-0.}---\phantom{0} $-0.05$ & \phantom{-0.}---\phantom{0}  \phantom{-0.}---\phantom{0}  & $+0.04$ $-0.08$  &$+0.04$  \phantom{-0.}---\phantom{0}   &  \phantom{-0.}---\phantom{0} \phantom{-0.}---\phantom{0}     \\
\noalign{\vskip 1mm}
\hline \hline
\end{tabular}
\label{tab:lfvsys}
\end{table*}

\begin{table*}
\centering
\caption{Systematic uncertainties from signal angular efficiency modeling. ``---'' denotes
where there is no uncertainty associated with a particular systematic.}
\begin{tabular}{l @{\extracolsep{1em}} c @{\extracolsep{1em}} c @{\extracolsep{1em}} c @{\extracolsep{1em}} c @{\extracolsep{1em}} c @{\extracolsep{1em}} c}
\\ \hline \hline \\
        & \multicolumn{3}{c}{$\fl$ systematic}         & \multicolumn{3}{c}{$\afb$ systematic} \\
\noalign{\vskip 1mm}         & $\modekstksll$    & $\modekstzll$     & $\modekstll$      & $\modekstksll$    & $\modekstzll$     & $\modekstll$      \\ \hline
\noalign{\vskip 1mm} $q^2_0$ &  \phantom{-0.}---\phantom{0}  $-0.02$    &  \phantom{-0.}---\phantom{0}   \phantom{-0.}---\phantom{0}      &  \phantom{-0.}---\phantom{0}   \phantom{-0.}---\phantom{0}      & $+0.04$  \phantom{-0.}---\phantom{0}      & $+0.04$  \phantom{-0.}---\phantom{0}      & $+0.05$  \phantom{-0.}---\phantom{0}      \\
\noalign{\vskip 1mm} $q^2_1$ & $+0.02$ $-0.04$    & $+0.14$  \phantom{-0.}---\phantom{0}      & $+0.13$  \phantom{-0.}---\phantom{0}      &  \phantom{-0.}---\phantom{0}  $-0.13$     &  \phantom{-0.}---\phantom{0}  $-0.20$     &  \phantom{-0.}---\phantom{0}  $-0.17$     \\
\noalign{\vskip 1mm} $q^2_2$ &  \phantom{-0.}---\phantom{0}  $-0.07$    &  \phantom{-0.}---\phantom{0}  $-0.10$     &  \phantom{-0.}---\phantom{0}  $-0.02$     & $+0.12$  \phantom{-0.}---\phantom{0}      & $+0.09$  \phantom{-0.}---\phantom{0}      & $+0.07$  \phantom{-0.}---\phantom{0}      \\
\noalign{\vskip 1mm} $q^2_3$ &  \phantom{-0.}---\phantom{0}  $-0.04$    & $+0.04$ $-0.05$     & $+0.02$ $-0.04$     & $+0.08$  \phantom{-0.}---\phantom{0}      & $+0.06$ $-0.04$     & $+0.07$ $-0.02$     \\
\noalign{\vskip 1mm} $q^2_4$ & $+0.07$ $-0.05$    & $+0.06$  \phantom{-0.}---\phantom{0}      & $+0.07$  \phantom{-0.}---\phantom{0}      & $+0.06$ \phantom{-0.}---\phantom{0}     &  \phantom{-0.}---\phantom{0}  $-0.09$     & $+0.02$ $-0.06$     \\
\noalign{\vskip 1mm} $q^2_5$ & $+0.10$  \phantom{-0.}---\phantom{0}     & $+0.02$  \phantom{-0.}---\phantom{0}      & $+0.07$  \phantom{-0.}---\phantom{0}      &  \phantom{-0.}---\phantom{0}  $-0.09$     &  \phantom{-0.}---\phantom{0}  $-0.08$     &  \phantom{-0.}---\phantom{0}  $-0.10$     \\
\noalign{\vskip 1mm}
\hline \hline
\end{tabular}
\label{tab:genvarsys}
\end{table*}

\begin{table*}
\centering
\caption{Total systematic uncertainties.}
\begin{tabular}{l @{\extracolsep{1em}} c @{\extracolsep{1em}} c @{\extracolsep{1em}} c @{\extracolsep{1em}} c @{\extracolsep{1em}} c @{\extracolsep{1em}} c}
\\ \hline \hline \\
        & \multicolumn{3}{c}{$\fl$ systematic}         & \multicolumn{3}{c}{$\afb$ systematic} \\
\noalign{\vskip 1mm}         & $\modekstksll$    & $\modekstzll$     & $\modekstll$      & $\modekstksll$    & $\modekstzll$     & $\modekstll$     \\ \hline
\noalign{\vskip 1mm} $q^2_0$ & $+0.02$  $-0.10$  & $+0.02$  $-0.02$  & $+0.02$  $-0.02$  & $+0.08$  $-0.05$  & $+0.06$  $-0.05$  & $+0.07$  $-0.09$ \\
\noalign{\vskip 1mm} $q^2_1$ & $+0.09$  $-0.14$  & $+0.15$  $-0.02$  & $+0.13$  $-0.05$  & $+0.13$  $-0.16$  & $+0.10$  $-0.21$  & $+0.08$  $-0.19$ \\
\noalign{\vskip 1mm} $q^2_2$ & $+0.18$  $-0.10$  & $+0.02$  $-0.10$  & $+0.02$  $-0.02$  & $+0.36$  $-0.49$  & $+0.12$  $-0.11$  & $+0.14$  $-0.11$ \\
\noalign{\vskip 1mm} $q^2_3$ & $+0.05$  $-0.08$  & $+0.05$  $-0.05$  & $+0.05$  $-0.07$  & $+0.08$  $-0.20$  & $+0.08$  $-0.08$  & $+0.08$  $-0.05$ \\
\noalign{\vskip 1mm} $q^2_4$ & $+0.16$  $-0.15$  & $+0.06$  $-0.06$  & $+0.07$  $-0.10$  & $+0.11$  $-0.24$  & $+0.17$  $-0.16$  & $+0.14$  $-0.13$ \\
\noalign{\vskip 1mm} $q^2_5$ & $+0.10$  $-0.21$  & $+0.02$  $-0.11$  & $+0.07$  $-0.14$  & $+0.18$  $-0.17$  & $+0.10$  $-0.10$  & $+0.10$  $-0.12$ \\
\noalign{\vskip 1mm}
\hline \hline
\end{tabular}
\label{tab:totalsys}
\end{table*}

\begin{table*}
\centering
\caption[$P_2$ results with total uncertainties.]
{$P_2$ results with total uncertainties.}
\begin{tabular}{l @{\extracolsep{4em}} c @{\extracolsep{4em}} c @{\extracolsep{4em}} c}
\\ \hline \hline \\
                      & $\modekstksll$          & $\modekstzll$           & $\modekstll$ \\ \hline \noalign{\vskip 1mm}
$q^2_0$               & $-0.22_{-0.13}^{+0.14}$ & $-0.07_{-0.21}^{+0.20}$ & $-0.18_{-0.13}^{+0.13}$ \\ \noalign{\vskip 1mm}
$q^2_1$               & $-0.29_{-0.17}^{+0.19}$ & $+0.12_{-0.29}^{+0.27}$ & $-0.09_{-0.17}^{+0.18}$ \\ \noalign{\vskip 1mm}
$q^2_2$               & $-0.38_{-0.28}^{+0.35}$ & $-0.27_{-0.24}^{+0.25}$ & $-0.35_{-0.16}^{+0.19}$ \\ \noalign{\vskip 1mm}
$q^2_3$               & $-0.09_{-0.21}^{+0.24}$ & $-0.22_{-0.22}^{+0.27}$ & $-0.14_{-0.13}^{+0.15}$ \\ \noalign{\vskip 1mm}
$q^2_4$               & $-0.15_{-0.26}^{+0.28}$ & $-0.48_{-0.27}^{+0.34}$ & $-0.42_{-0.20}^{+0.26}$ \\ \noalign{\vskip 1mm}
$q^2_5$               & $-0.95_{-0.96}^{+1.84}$ & $-0.37_{-0.24}^{+0.28}$ & $-0.41_{-0.21}^{+0.34}$ \\ \noalign{\vskip 1mm}
\hline \hline
\label{tab:p2results}
\end{tabular}
\end{table*}

\subsection{Extraction of $P_2$ from the Angular Fit Results}

\begin{figure*}[]
\begin{center}
\includegraphics[width=.95\textwidth]{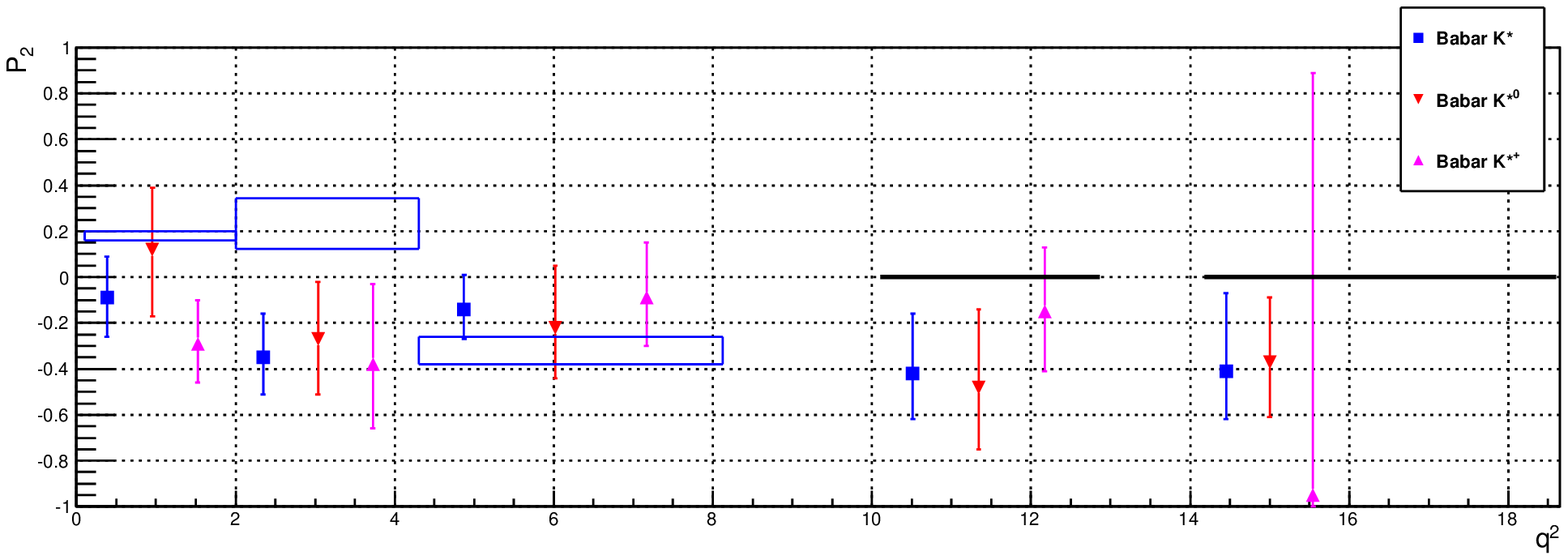}
\caption{$P_2$ results with total uncertainties.The blue boxes show
the SM theory expectation in the low mass-squared region; there are
no comparable calculations in the high mass-squared region, where the
black lines simply denote the extent of the $q^2_4$ and $q^2_5$ bins.}
\label{fig:p2results}
\end{center}
\end{figure*}

As mentioned above in the Introduction, $\fl$ and $\afb$
can be used to parameterize an additional angular observable,
$P_2 = (-2/3) * \afb / (1-\fl)$, which has diminished theory
uncertainty and greater sensitivity to non-SM contributions
than either $\fl$ or $\afb$ alone~\cite{NewPhysicsP2,dc}. Table~\ref{tab:p2results}
gives our results for $P_2$. The 68\% confidence intervals quoted are
frequentist and derived from ensembles of fits
to simulated datasets randomly drawn from the correlated
confidence-level contours for $\fl$ and $\afb$.
For the disjoint mass-squared bins $q^2_1$ to $q^2_5$,
Fig.~\ref{fig:p2results} graphically shows our results overlaid
on the SM expectations from theory (as given in Table~3 (KMPW) of
Ref.~\cite{dc}) in the mass-squared region below the $\jpsi$.
In the $q^2_0$ mass-squared bin, the SM expectation (from the
same source) for $P_2$ is $0.11 \pm 0.10$, in slight tension
with our experimental result.

\section{Conclusion}
In conclusion, we have measured in bins of dilepton mass-squared
the fraction $\fl$ of longitudinally polarized $\Kstar$ decays
and the lepton forward-backward asymmetry $\afb$
in the decays $\modekstksll$, $\modekstzll$ and $\modekstll$.
Results for the $\modekstksll$ final state are presented for the first time here.
Fig.~\ref{fig:allresults} graphically shows our $\fl$ and $\afb$ results in disjoint $q^2$ bins
alongside other published results and the SM theory expectations, the latter of which
typically have 5-10\% theory uncertainties in the regions below and above
the charmonium resonances.
Fig.~\ref{fig:q0results} similarly compares the $q^2_0$ results obtained here with those of other
experiments and the SM theory expectation.
As shown in these figures, our $\modekstzll$ results are in reasonable agreement with both
SM theory expectations and other experimental results.
Similarly, although with relatively larger uncertainties, we observe broad agreement of
the $\modekstksll$ results with those for $\modekstzll$.
However, in the low dilepton mass-squared region, we observe relatively very
small values for $\fl$ in $\modekstksll$, exhibiting tension
with both the $\modekstzll$ results as well as the SM expectations.
Similarly, as shown in Fig.~\ref{fig:p2results} in the same mass-squared region, there appears to be some
tension between the experimental results and the expected SM values
for $P_2$.

\section{ACKNOWLEDGMENTS}
We would like to express our gratitude to Joaquim Matias for discussions
regarding the observable $P_2$. We are grateful for the
extraordinary contributions of our \pep2\ colleagues in achieving
the excellent luminosity and machine conditions that have made this
work possible. The success of this project also relies critically on the
expertise and dedication of the computing organizations that support \babar. The
collaborating institutions wish to thank SLAC for its support and the kind
hospitality extended to them. This work is supported by the US Department of
Energy and National Science Foundation, the Natural Sciences and Engineering
Research Council (Canada), the Commissariat \`a l'Energie Atomique and Institut
National de Physique Nucl\'eaire et de Physique des Particules (France), the
Bundesministerium f\"ur Bildung und Forschung and Deutsche
Forschungsgemeinschaft (Germany), the Istituto Nazionale di Fisica Nucleare
(Italy), the Foundation for Fundamental Research on Matter (The Netherlands),
the Research Council of Norway, the Ministry of Education and Science of the
Russian Federation, Ministerio de Econom\'{\i}a y Competitividad (Spain), the
Science and Technology Facilities Council (United Kingdom), and the Binational
Science Foundation (U.S.-Israel). Individuals have received support from the
Marie-Curie IEF program (European Union) and the A. P. Sloan Foundation (USA).

\end{document}